\documentclass[11pt,a4paper]{article} 
\usepackage{amssymb,bbm}
\usepackage[all]{xy}
\font\got=eufm10 at 12pt
\newtheorem{theorem}{Theorem}[section] 
\newtheorem{lemma}[theorem]{Lemma}
\newtheorem{proposition}[theorem]{Proposition}
\newtheorem{corollary}[theorem]{Corollary}
\newtheorem{definition}[theorem]{Definition}
\newtheorem{remark}[theorem]{Remark}
\newtheorem{example}[theorem]{Example}

\def\1{{\mathbbm{1}}}
\def\A{{\cal A}}
\def\Ad{{\mathop{\rm Ad}}}
\def\be{\begin{equation}}
\def\build#1_#2^#3{\mathrel{\mathop{\kern 0pt#1}\limits_{#2}^{#3}}}
\def\C{{\mathcal {C}}}
\def\codim{{\mathop{\rm codim \;}}}
\def\E{{\mathbb{E}}}
\def\EC{{\wG^{(\E)}}}

\def\ee{\end{equation}}

\def\epsilon{\varepsilon}
\def\F{{\mathbb{F}}}
\def\G{{\mathbb{G}}}

\def\id{{\rm{id}}}
\def\LG{{\hbox{\got g}}}
\def\J{{\mathcal{J}}}
\def\lra{\longrightarrow}
\def\M{{\cal M}}
\def\N{{\cal N}}

\def\oo{{\hbox{\got{o}}}}
\def\P{{\mathbb{P}}}
\def\pf{\noindent \textbf{Proof -- }}
\def\Q{{\mathbb{Q}}}
\def\qed{\hfill\hbox{\vrule\vbox to 2mm{\hrule width 2mm\vfill\hrule}\vrule}
  \\}
\def\R{{\mathbb{R}}}
\def\S{{\cal S}}
\def\T{{\cal {T}}}

\def\V{{\mathbb{V}}}

\def\wG{{\widetilde{G}}}
\def\wJ{\widetilde{\mathcal{J}}}

\def\wS{\widetilde{S}}
\def\wT{T}
\def\xx{{\hbox{\got {x}}}}
\def\Z{{\mathbb{Z}}}

\title{Discrete and continuous Yang-Mills measure for non-trivial
bundles over compact surfaces}
\author{Thierry L{\textsc{\'evy}}\thanks{CNRS, DMA - ENS - 45, rue d'Ulm
- 75005 Paris}}

\begin{document}

\maketitle

\begin{abstract}
In this paper, we construct one Yang-Mills measure on a compact
surface for each isomorphism class of principal bundles over this
surface. For this, we refine the discretization procedure used in a
previous construction \cite{Levy_AMS} and define a new discrete theory
which is essentially a covering of the usual one. We prove that the
measures corresponding to different isomorphism classes of bundles or
to different total areas of the base space are mutually singular. We
give also a combinatorial computation of the partition functions which
relies on the formalism of fat graphs. 
\end{abstract}

\section*{Introduction}

The Yang-Mills measure is the law of a group-valued random process
indexed by a family of paths on some manifold. It is usually though of
as the random holonomy induced by a probability measure on the space
of connections on a principal bundle over this manifold. We consider
in this paper the case where the base manifold is an oriented compact
surface and the structure group is a compact connected Lie group. In
this case, the measure has been studied at various levels of rigor by
several authors.  In particular, the origin of its mathematical study
is a paper by the physicist A. Migdal \cite{Migdal}. Other important
contributions are those of B. Driver \cite{Driver,Driver_lassos} and,
with different motivations, E. Witten \cite{Witten}. The first
rigorous construction has been given by A. Sengupta
\cite{Sengupta_AMS}, by conditioning an infinite-dimensional noise. A
second construction has been given by the author in \cite{Levy_AMS},
where the random holonomy process is built by passing discrete
approximations to the limit. This leads essentially to the same
object, though perhaps in a way that gives a better grip on it (see
for example \cite{Levy_Norris}).
 
An important feature of the Yang-Mills measure in two dimensions is
its almost invariance by diffeomorphisms (it is actually invariant by
those which preserve a volume form on the surface). Still, a
diffeomorphism-invariant random holonomy process on a principal bundle
should depend not only on the structure group of this bundle but also
on its isomorphism class. This is especially clear if one thinks that
the characteristic classes of a bundle can be computed using
connections on it. However, the construction proposed in
\cite{Levy_AMS} does not depend on the particular topological type of
the principal bundle one considers. We shall see that the measure
constructed there corresponds to an average over all possible
isomorphism classes. On the other hand, A. Sengupta's construction
does allow one to take a specific topological type of bundle into
account. The aim of this paper is to fill this gap. More precisely, we
propose a construction of the Yang-Mills measure by passing a discrete
approximation to the continuous limit {\it in a way that keeps track
of the topology of the bundle}.
 
This may sound paradoxical for the following reason.  Discrete
approximations of the measures are usually built by first considering
graphs on the base manifold and restricting the bundle to these
one-dimensional complexes. But, as long as the structure group is
connected, the restricted bundle is always trivial and any topological
information about the full bundle is lost in this operation. This is
why the construction of \cite{Levy_AMS} produces a single probability
measure, which is associated to no particular isomorphism class of
fiber bundles, but rather to a random bundle for some natural measure
on the set of isomorphism classes. In the case of an Abelian structure
group, this was to some extent already understood, because in this
case it is easier to compare A. Sengupta's construction and ours (see
the informal remarks in \cite[Sections 1.9.3 and 3.2]{Levy_AMS}).
 
The first section of this paper is devoted to the construction of a
new discrete theory, namely a finer way of discretizing the
differential geometric objects involved than the naive one commonly
used in discrete gauge theory. This new discretization is fine enough
to capture the topology of the bundle. In fact, what we do is
replacing the usual configuration space of discrete gauge theory by a
singular covering of it, which happens to be finite when the structure
group is semi-simple. In all cases, the singular set is negligible and
plays no role at the level of measure theory. A partial study of the
topological structure of this singular covering is presented in the
last section.

In the second and third sections, we construct the Yang-Mills measures
associated to a specific isomorphism class of principal bundles. For
this, we consider first semi-simple structure groups, because in this
case, as explained above, the new configuration space is a singular
{\it finite} covering of the usual one. Then, although the method used
to construct the discrete measures does not extend to general
structure groups, the formul\ae{} derived in the semi-simple case make
sense without the semi-simplicity assumption. This allows us to
construct the discrete measures for general compact connected
structure groups. In the case of Abelian groups, we check that our
construction is consistent with the remarks of \cite{Levy_AMS}
mentioned above. Finally, in the third section, we pass these discrete
measures to the continuous limit, following step by step the
construction presented in the second chapter of
\cite{Levy_AMS}. However, balancing the fact that it carries more
geometric informations, the new discretization does not produce a nice
projective family of probability spaces as the usual procedure
does. It is thus necessary to go back to the usual configuration
spaces before passing to the limit. As a consequence, we construct
probability measures on the same space as in \cite{Levy_AMS}, not on
some covering of it.

At the end of the third section, we prove that the Yang-Mills measures
corresponding to different isomorphism classes of principal bundles or
to different temperatures (one should say, different total areas of
the base manifold) are mutually singular. This shows that the
canonical probability space of the random holonomy process is a
disjoint union of infinitely many sectors, each one corresponding to a
specific isomorphism class of bundles and a specific temperature.
 
The fact that the discrete partition functions do not depend on the
graph one considers plays an important role. The proof of this fact
given in \cite{Levy_AMS} relies on rather painful approximations. We
present in the fourth section a much more natural proof of this
invariance, based on the formalism of fat graphs, which seems very
well suited to discrete gauge theory. For example, it enables us to
compute directly these partition functions in a quite intuitive
combinatorial way.\\

{\bf Acknowledgement -- } Part of this work was initiated in 2001 when
the author was enjoying the kind hospitality of the Statistical
Laboratory in Cambridge (UK).

\section{The configuration space}
\label{section:config}
Let $M$ be a compact oriented surface without boundary. Let $\sigma$
be a volume $2$-form on $M$ consistent with the orientation. For
practical purposes, let us endow $M$ with a Riemannian metric and let
us assume that the corresponding Riemannian volume is the density of
$\sigma$. Let $G$ be a compact connected Lie group. Let $P \lra M$ be a
principal $G$-bundle over $M$. 

If $C$ is a closed subset of $M$, we call smooth mapping
(resp. smooth cross-section, $\ldots$) on $C$ the restriction to $C$ of a
smooth mapping (resp. smooth cross-section, $\ldots$) defined on an open
neighbourhood of $C$. In particular, we denote by $\Gamma(C,P)$ the
set of smooth cross-sections of $P$ over $C$.

\subsection{Graphs}
\label{sect: graphs}

By an {\it edge} on $M$ we mean a segment of a smooth oriented
1-dimensional submanifold. If $e$ is an edge, we call {\it inverse} of
$e$ and denote by $e^{-1}$ the edge obtained by reversing the
orientation of $e$. We also denote respectively by $\underline e$ and
$\overline e$ the starting and finishing point of $e$. Let
$e_1,\ldots,e_n$ be $n$ edges. If, for each $i$ between $1$ and $n-1$,
one has $\overline{e_i}=\underline{e_{i+1}}$, then one can form the
concatenation $e_1\ldots e_n$. If moreover $f_1,\ldots,f_m$ are also
edges which can be concatenated, we declare $e_1\ldots e_n$ equivalent
to $f_1\ldots f_m$ if and only if there exists a continuous mapping
$c:[0,1] \lra M$ and two finite sequences $0=t_0<t_1<\ldots<t_n=1$ and
$0=s_0<s_1<\ldots < s_m=1$ of real numbers such that, for each
$i=1\ldots n$, the restriction of $c$ to the interval $[t_{i-1},t_i]$
is a smooth embedding of image $e_i$ and, for each $j=1\ldots m$, the
restriction of $c$ to the interval $[t_{j-1},t_j]$ is a smooth
embedding of image $f_j$. By a {\it path} we mean an equivalence class
of finite concatenations of edges. We denote the set of paths by
$PM$. Loops, simple loops, starting and finishing points of paths,
their concatenation, are defined in the obvious way. Let $l_1$ and
$l_2$ be two loops. We say that $l_1$ and $l_2$ are cyclically
equivalent if there exist two paths $c$ and $d$ in $PM$ such that $l_1
= cd$ and $l_2=dc$. We call {\it cycle} an equivalence class of loops
for this relation. Informally, a cycle is a loop on which one has
forgotten the starting point. We say that a cycle is {\it simple} if
its representatives are simple loops.

\begin{definition}
A {\sl graph} is a triple $\G=(\V,\E,\F)$, where

1. $\E$ is a finite collection of edges stable by inversion and such
   that two distinct edges are either inverse of each other or
   intersect, if at all, only at some of their endpoints.

2. $\V$ is the set of endpoints of the elements of $\E$.

3. $\F$ is the set of the closures of the connected components of $M
   \backslash \bigcup_{e\in\E} e$.

4. Each open connected component of $M \backslash \bigcup_{e\in\E} e$
   is diffeomorphic to the open unit disk of $\R^2$.
\end{definition}

The elements of $\V,\E,\F$ are respectively called vertices, edges and
faces of $\G$. We call {\it open faces} the connected components of $M
\backslash \bigcup_{e\in\E} e$. Beware that an open face may be
strictly contained in the interior of its closure. 

We denote by $\E^*$ the set of paths that can be represented by a
concatenation of elements of $\E$.

This definition of a graph is that of \cite{Levy_AMS}. In the first
part of this paper, we are going to work with a slightly more
restrictive notion of a graph. Before defining it, let us explain what
we mean by the boundary of a face.

Let $\G$ be a graph. The orientation of $M$ determines, at each
vertex, a cyclic order on the set of incoming edges. It is defined as
the order of the intersection points of these edges with a small
geodesic circle around the vertex. This means that $\G$ induces a
structure of {\it fat graph} on $\E$ (see \cite{Lando_Zvonkin} or
Section \ref{fat graphs} for a presentation of fat graphs). This fat
graph has faces, which are defined as the cycles of some permutation
on $\E$. Each such cycle is of the form $(e_1,\ldots,e_m)$ with
$\overline{e_i}=\underline{e_{i+1}}$ for all $i=1\ldots m-1$ and
$\overline{e_m}=\underline{e_1}$. Hence, it defines a cycle in $\E^*$.

Let now $L: \E \lra \F$ be the mapping defined by the fact that, for
each edge $e$, $L(e)$ is the face of $\G$ located on the left of $e$,
that is, the closure of the unique open face of $\G$ which $e$ bounds
with positive orientation. The function $L$ is constant on the cycles
of edges corresponding to the faces of the fat graph induced by $\G$
and, since the open faces are diffeomorphic to disks, this sets up a
one-to-one correspondence between the faces of the fat graph induced
by $\G$ and the elements of $\F$. For each $F\in \F$, the cycle
associated in this way with $F$ is called the boundary of $F$ and it
is denoted by $\partial F$. Recall that its origin is
ill-defined. However, the following definition makes sense.

\begin{definition} We say that a graph is {\sl simple} if the boundary
of each one of its faces is a simple cycle.
\end{definition}

Let $\G$ be a graph. We call {\it unoriented edge} of $\G$ a pair
$\{e,e^{-1}\}$ where $e \in \E$. We call {\it orientation} of $\G$ a
subset $\E^+$ of $\E$ which contains exactly one element of each
unoriented edge. 

\subsection{Discretization of a connection on $P$}
\label{discretisation connexion}

Let $\omega$ be a connection $1$-form on $P$. Let $\G=(\V,\E,\F)$
be a simple graph on $M$. 

Let $s_\V \in \Gamma(\V,P)$ be a cross-section of $P$ over $\V$. Let $F$ be
a face of $\G$. Since the group $G$ is connected, there exists a
smooth cross-section $s_F \in \Gamma(F,P)$ of $P$ over $F$ such that
${s_F}_{|F \cap \V} = {s_\V}_{|F\cap \V}$. Let us choose such a
section for each $F$ and set $s=(s_F)_{F\in \F}$. This is an element
of the set $\S(\G,P)$ defined by
$$\S(\G,P)=\{s \in \prod_{F\in \F} \Gamma(F,P) \; | \; \forall
F,F' \in \F, {s_F}_{|F\cap F' \cap \V} = {s_{F'}}_{|F \cap F' \cap
\V}\}.$$

Let $\pi: \wG \lra G$ be a universal cover of $G$. We shall explain
now how the choice of $s$ allows us to associate one element of $\wG$
to each edge of $\G$. Let $\LG$ be the Lie algebra of $G$. The
covering map determines an isomorphism of Lie algebras through which
we identify $\LG$ with the Lie algebra of $\wG$.

Take $e$ in $\E$. Set $F=L(e)$. Let us parametrize $e$ smoothly as
$e:[0,1] \lra M$. Now the ordinary differential equation 
\be
\label{holonomie} \cases{h_0=1 \cr \dot h_t h_t^{-1} = -(s_F^*
\omega)(\dot e_t) \;\; ,\;\; t\in [0,1]} 
\ee 
which one usually solves in $C^\infty([0,1],G)$ in order to determine
the holonomy of $\omega$ along $e$ can just as well be solved in
$C^\infty([0,1],\wG)$. This is what we do and we denote by $\tilde
g(e)$ the element $h_1$ of $\wG$.

\begin{lemma}\label{meme projection}
For all $e\in \E$, one has $\pi(\tilde g(e))= \pi(\tilde
g(e^{-1}))^{-1}$. 
\end{lemma}

\pf The function $t\mapsto \pi(h_t)$ is the solution of
(\ref{holonomie}) in $C^\infty([0,1],G)$. Hence, $\pi(\tilde g(e))$
and $\pi(\tilde g(e^{-1}))^{-1}$ are both equal to the unique element
$x$ of $G$ such that the $\omega$-horizontal lift of $e$ starting at
$s_\V(\underline e)$ ends at $s_\V(\overline e)x$. \qed

Thus, the choice of $s$ allows us to define an element of what we take
as the new configuration space for the discrete Yang-Mills theory on
$\G$ with structure group $G$, namely
$$\wG^{(\E)}=\{ \tilde g \in \wG^\E \; | \; \forall e \in \E,
\pi(\tilde g(e))= \pi(\tilde g(e^{-1}))^{-1}\}.$$ More precisely, if
$\A$ denotes the set of smooth\footnote{It is enough to consider
connections in the Sobolev space $H^1$ in order to be able to solve
(\ref{holonomie}) and define a holonomy, see \cite{Levy_Norris} for
more details.} connection $1$-forms on $P$, we have defined a mapping
\begin{eqnarray}
\tilde H_\G:\S(\G,P) \times \A & \lra & \wG^{(\E)} \nonumber \\
(s,\omega) & \longmapsto & \tilde H_\G(s,\omega)=(\tilde
g^s_\omega(e))_{e\in \E} \label{*}
\end{eqnarray}

\subsection{Gauge transformations}

Let $\Pi \subset \wG$ be the kernel of $\pi : \wG \lra G$. It is a
discrete central subgroup of $\wG$ and it is finite if and only if $G$
is semi-simple. We use the generic notation $z$ for the elements of
$\Pi$, reminding us in this way that they are central.

Consider the homomorphism
\begin{eqnarray*}
\Pi^\E & \lra & \Pi^\F \\
(z_e)_{e\in \E} & \longmapsto & \left( \prod_{L(e)=F} z_e
\right)_{F\in \F}
\end{eqnarray*}
and let $J_\Pi$ denote its kernel. 

\begin{definition} We call {\sl discrete gauge group} the group $\wJ_\G=
\wG^\V \times J_\Pi$. This group acts on $\wG^{(\E)}$ as follows: given $\tilde
g=(\tilde g(e))_{e\in \E} \in \wG^{(\E)}$ and $j =((j_v)_{v\in \V},
(z_e)_{e\in \E}) \in \wJ_\G$, $j \cdot \tilde g$ is defined by
\be \label{action jauge}
\forall e\in \E \; , \;\; (j\cdot \tilde g)(e)=j_{\overline e}^{-1} \tilde g(e) j_{\underline
e} z_e.
\ee
\end{definition}

Let $Z(\wG)$ denote the center of $\wG$. Consider the homomorphism 
\begin{eqnarray*}
Z(\wG) \times \Pi^\V & \lra & \wJ_\G \\
(\tilde x, (k_v)_{v\in \V}) & \longmapsto & ((\tilde x
k_v)_{v\in\V},(k_{\overline{e}} k_{\underline{e}}^{-1})_{e\in\E})
\end{eqnarray*}
and let ${\cal K}_\G$ denote its image.

\begin{proposition} \label{fidele}
The group $\J_\G$ acts on $\wG^{(\E)}$ with kernel ${\cal K}_\G$ and
its orbits satisfy the following property: if $s$ belongs to
$\S(\G,P)$ and $\omega$ to $\A$, then
\be \label{3}
\bigcup_{s' \in \S(\G,P)} \{\tilde H_\G(s',\omega)\} = \wJ_\G \cdot \tilde
H_\G(s,\omega).
\ee
\end{proposition}

\pf That the kernel of the action is ${\cal K}_\G$ follows easily from
the definitions of the action and ${\cal K}_\G$. In order to prove
(\ref{3}), fix $s$ in $\S(\G,P)$ and $\omega$ in $\A$.  Take $s'$ in
$\S(\G,P)$. There exists $u \in \S(\G,M\times G)$ such that $s'=su$,
that is, for each $F$, $s'_F=s_F u_F$. Since $\G$ is a simple graph,
its faces are contractible and it is possible for each $F\in \F$ to
lift $u_F$ to a mapping $\tilde u_F : F \lra \wG$.  Observe however
that this does not define an element of $\S(\G,M\times \wG)$ as there
is no guarantee that the lifts coincide over the vertices of the
graph. Nevertheless, these lifts allow us to express $\tilde
H_\G(s',\omega)=\tilde g^{s'}_\omega$ in function of $\tilde
H_\G(s,\omega)=g^s_\omega$. Indeed, for each $e\in \E$, \be
\label{4} \tilde g^{s'}_\omega(e)= \tilde u_{L(e)}(\overline e)^{-1}
\tilde g^s_\omega(e) \tilde u_{L(e)}(\underline e), \ee which does not
depend on the choice of the lift.

Let $u_\V$ be the cross-section of $M\times G$ over $\V$ determined by
$u$. Let $\tilde u_\V$ be a lift of $u_\V$ to a section of $M\times
\wG$. Then (\ref{4}) can be rewritten as
$$\tilde g^{s'}_\omega(e)= \tilde u_\V(\overline e)^{-1} \tilde
g^s_\omega(e) \tilde u_\V(\underline e) \left[\tilde
u_{L(e)}(\overline e)^{-1} \tilde u_\V(\overline e)\right]
\left[\tilde u_{L(e)}(\underline e)^{-1} \tilde u_\V(\underline
e)\right]^{-1}.$$

Set, for each $v\in \V$, $j_v=\tilde u_\V(v)$ and, for each $e\in \E$,
$$z_e=\left[\tilde u_{L(e)}(\overline e)^{-1}\tilde u_\V(\overline
e)\right] \left[\tilde u_{L(e)}(\underline e)^{-1} \tilde
u_\V(\underline e)\right]^{-1}.$$
Then, $j=((j_v)_{v\in\V}, (z_e)_{e\in \E})$ belongs to $\wJ_\G$ and
satisfies $j\cdot \tilde H(s,\omega)=\tilde H(s',\omega).$

Now let $j=((j_v)_{v\in\V}, (z_e)_{e\in \E})$ be an element of
$\wJ_\G$. Let $\tilde u_\V \in \Gamma(\V,M\times \wG)$ be defined by
$\tilde u_\V(v)=j_v$. Let $F$ be a face of the graph. Let $e_1\ldots
e_n$ be a simple loop which represents $\partial F$. We construct a
cross-section $\tilde u_F$ of $M\times \wG$ over $F$. The conditions
$\tilde u_F(\underline e_1)=\tilde u_\V(\underline e_1)$ and
$$z_{e_i}=\left[\tilde u_{F}(\overline {e_i})^{-1}\tilde
u_\V(\overline {e_i})\right] \left[\tilde u_{F}(\underline
{e_i})^{-1} \tilde u_\V(\underline {e_i})\right]^{-1}$$ for each
$i=1\ldots n$ determine the values of $\tilde u_F$ over the vertices
located on the boundary of $F$. Observe that $\tilde
u_F(\underline{e_1})$ is well-defined because $e_1\ldots e_n$ is a
simple loop and $z_{e_1} \ldots z_{e_n}=1$. Now, since $\wG$ is
connected, $\tilde u_F$ can be extended to the boundary of $F$ and
even, since $\wG$ is simply connected, to $F$ itself. Let $u_F$ be the
projection on $G$ of $\tilde u_F$. The value at a vertex
$v$ of $u_F$ is $\pi(j_v)$.

Doing this for each face produces an element $u \in \S(\G,M\times
G)$. It follows now from (\ref{4}) that this element satisfies $\tilde
H(su,\omega)= j \cdot \tilde H(s,\omega)$. \qed

Let $\J$ denote the smooth gauge group\footnote{If one considers $H^1$
connections rather than smooth ones, then one should consider $H^2$
gauge transformations, see \cite{Uhlenbeck,Wehrheim}.} of $P$. It acts
on $\A$ by pull-back.

\begin{corollary} \label{action j}
Let $j \in \J$ and $\omega \in \A$. Then, for all
$s\in \S(\G,P)$, one has $\tilde H_\G(s,j\cdot \omega) \in \wJ_\G \cdot
\tilde H_\G(s,\omega)$.
\end{corollary}

\pf Since $\tilde H_\G(s,j\cdot \omega)$ is built from pull-backs of
$j\cdot \omega$ by $s$, which are the same as the pull-backs of
$\omega$ by $j(s)$, one has $\tilde H_\G(s,j\cdot \omega)=\tilde
H_\G(j(s),\omega)$ and the result follows. \qed

Finally, Proposition \ref{fidele} and Corollary \ref{action j} show
that (\ref{*}) induces a mapping $\tilde H_\G: \A/\J \lra \EC/\wJ_\G$.

\subsection{Comparison with the classical formalism of discrete gauge theory}

Let $\E^+$ be an orientation of $\G$. The usual configuration space in
discrete Yang-Mills theory is $G^{\E^+}$. There is a mapping
$$H_\G : \Gamma(\V,P) \times \A \lra G^{\E^+}$$
defined as follows. Pick $s_\V \in \Gamma(\V,P)$ and $\omega \in
\A$. Then $H_\G(s_\V,\omega)=(g(e))_{e\in \E^+}$, where, for each $e\in
\E^+$, the $\omega$-horizontal lift of $e$ starting at
$s_\V(\underline{e})$ finishes at $s_\V(\overline{e})g(e)$.

The discrete gauge group in this setting is $\J_\G=G^\V/Z(G)$, where $Z(G)$
is embedded diagonally in $G^\V$. It acts faithfully on $G^{\E^+}$ and
satisfies a property similar to (\ref{3}). Hence, $H_\G$ induces a
mapping $\A \lra G^{\E^+}/\J_\G$ and even, by the same argument as
Corollary \ref{action j}, a mapping $H_\G: \A/\J \lra G^{\E^+}/\J_\G$.

Of course, this construction depends on the orientation $\E^+$, but
Lemma \ref{meme projection} ensures that $\pi : \wG \lra G$ induces a
covering $\pi : \EC \lra G^{\E^+}$ which is consistent with the choice
of orientation. The covering of $G$ induces also a covering map $\pi :
\wJ_\G \lra \J_\G$ in such a way that, for all $j \in \wJ_\G$, $\tilde
g \in \EC$, $\pi(j\cdot \tilde g) = \pi(j) \pi(\tilde g).$ This
equivariance property implies that $\pi : \EC \lra G^{\E^+}$ maps each
orbit of $\wJ_\G$ onto an orbit of $\J_\G$ and induces a mapping
between the topological quotient spaces $\pi : \EC/\wJ_\G \lra
G^{\E^+}/\J_\G$.
Finally, the following diagram commutes.
\be \label{lift H}
\centerline{\xymatrix{& \EC/\wJ_\G \ar[dd]^\pi \\\A/\J \ar[ru]^{\tilde
H_\G} \ar[rd]_{H_\G} &\\ & G^{\E^+}/\J_\G}}
\ee

We prove in the last section that this mapping $\pi$ is, outside a
negligible singular set, a covering with fiber isomorphic to $\Pi^\F$. 

%\be \label{lift H}
%\centerline{\xymatrix{& \EC/\wJ_\G \ar[dd]^\pi & \supset &  \EC/\wJ -
%\wN  \ar[r]^\kappa & (G^{\E^+}/\J - N)
%\times \Pi^\F \ar[ldd]_{p_1}  \\\A/\J \ar[ru]^{\tilde
%H_\G} \ar[rd]_{H_\G} &&&& \\ & G^{\E^+}/\J_\G & \supset & G^{\E^+}/\J
%- N  &}}
%\ee
The new discrete theory that we present in this paper is thus
essentially a lift of the usual discrete theory. We are now going to
show that this lift contains a geometric information which is not
present at the level of the usual discrete gauge theory.

\subsection{The obstruction class of the bundle}

Let us fix a connection $\omega\in \A$ and an element $s$ of
$\S(\G,P)$.

The principal $G$-bundle bundle $P$ is classified up to isomorphism by
a cohomology class of $H^2(M;\pi_1(G))$ (see \cite{Steenrod}). Using
the orientation of $M$, we identify this class with an element of
$\pi_1(M)$ which in turn we identify with an element of $\Pi$, which
we denote by $\oo(P)$. It turns out that, once a connection $\omega$
on $P$ is chosen, $\oo(P)$ can be extracted from $\tilde
H_\G(\omega)=(\tilde g(e))_{e\in\E}$ in a very simple way.

\begin{lemma}\label{obstruction}
Let $\E^+$ be an orientation of $\G$. The following equality holds:
\be \label{obs class}
\oo(P)=\prod_{e\in \E^+} \tilde g(e) \tilde g(e^{-1}).
\ee
\end{lemma}

\pf The element $\oo(P)$ of $\pi_1(G)$ can be defined as follows (see
\cite{Steenrod} and \cite{Morita} in the case $G=U(1)$). Choose an
element $s$ of $\S(\G,P)$. Let $e$ be an edge. There exists a unique
smooth mapping $\delta_e : e \lra G$ such that the equality
$s_{L(e^{-1})}= s_{L(e)} \delta$ holds identically over $e$. Actually,
$\delta_e$ maps $e$ to a loop in $G$ based at the unit element and
whose homotopy class is denoted by $[\delta_e]$. Then, $\prod_{e\in
\E} [\delta_e]$ does not depend on the choice of $s$ and it is denoted
by $\oo(P)$.

On the other hand, it is easy to check that, via the identification
$\pi_1(G)\simeq \Pi$, the homotopy class $[\delta_e]$ corresponds to
$\tilde g(e) \tilde g(e^{-1})$. The result follows. \qed

We denote by $\oo : \wG^{(\E)} \lra \Pi$ the mapping defined by the right
hand side of (\ref{obs class}), which in fact does not depend on
$\E^+$. 

\begin{lemma} \label{invariance obstruction}
The mapping $\oo : \wG^{(\E)} \lra \Pi$ is invariant under the action of
$\wJ_\G$. 
\end{lemma}

\pf Let $j=((j_v)_{v\in \V},(z_e)_{e\in \E})$ be an element of
$\wJ_\G$ and $\tilde g$ an element of $\EC$. Let $\E^+$ be an
orientation of $\G$. One has
\begin{eqnarray*}
\oo(j\cdot \tilde g) &=& \prod_{e\in \E^+}
\Ad(j_{\overline{e}}^{-1})(\tilde g(e) \tilde g(e^{-1})) z_e
z_{e^{-1}} \\
&=& \oo(\tilde g) \prod_{e\in \E} z_e \\ 
&=& \oo(\tilde g).
\end{eqnarray*}
The result is proved. \qed

According to this lemma, we may regard $\oo$ as a function on
$\EC/\wJ_\G$. 

\begin{corollary} \label{obstr}
Let $P$ be a principal $G$-bundle over $M$ and
$\omega$ a connection on $P$. Let $\G$ be a simple graph on $M$. Then
$$\oo(P)=\oo(\tilde H_\G(\omega)).$$
\end{corollary}

We finish by explaining what amount of information about a pair
$(P,\omega)$ is encoded in the class $\tilde H_\G(\omega) \in
\wG^{(\E)}/\wJ_\G$.

\begin{proposition} Let $P$ and $Q$ be two principal $G$-bundles over
$M$ equipped respectively with two connections $\omega$ and
$\eta$. Let $\G$ be a simple graph on $M$. Let $i: \cup_{e\in\E} e
\hookrightarrow M$ denote the inclusion map. The following
propositions are equivalent:

\noindent (i) $\tilde H_\G(\omega) = \tilde H_\G(\eta)$.\\
\noindent (ii) There exists a bundle isomorphism $\varphi: P \lra Q$ such
that $i^* \varphi^* \eta = i^* \omega$.
\end{proposition}

\pf $(i) \Rightarrow (ii)$ By Corollary \ref{obstr}, $\oo(P)=\oo(Q)$, so
that $P$ and $Q$ are isomorphic. We may thus assume that $P=Q$. Now,
$\omega$ and $\eta$ determine the same class in the usual
configuration space $G^{\E^+}/G^\V$ and the result follows by
\cite[Lemma 1.11]{Levy_AMS}. 

$(ii) \Rightarrow (i)$ Let us consider $\tilde H_\G$ as defined by
(\ref{*}). Let $s$ be an element of $\S(\G,P)$. Then the assumption
implies $\tilde H_\G(s,\omega)= \tilde H_\G(s,\varphi^* \eta)$. Now,
$\varphi(s) \in \S(\G,Q)$ and $\tilde H_\G(s,\varphi^* \eta)= \tilde
H_\G(\varphi(s),\eta)$. Hence, $\tilde H_\G(\omega)=\tilde
H_\G(\eta)$. \qed

\section{The discrete measures}

In the first section, we have built a singular covering of the usual
configuration space of discrete Yang-Mills theory. We have explained
how the covering space $\EC$ is partitioned into several sectors,
each of which corresponds to an isomorphism class of principal
$G$-bundles: $\EC=\bigsqcup_{z\in\Pi} \oo^{-1}(z)$. 

In order to associate a probability measure on $G^{\E^+}$ to each
element $z$ of $\Pi$, we proceed as follows. First, we construct a
lift on $\EC$ of the usual discrete Yang-Mills measure on
$G^{\E^+}$. Then, for each $z$, we renormalize and project on
$G^{\E^+}$ the restriction of this lift to $\oo^{-1}(z)$. 

However, this strategy breaks down when $\wG \lra G$ is not a finite
covering, that is, when $G$ is not semi-simple. In fact, in this case,
the lift of the discrete measure does not exist as a probability
measure, but as an infinite measure on $\EC$ and it is not possible to
renormalize properly its restriction to a given sector
$\oo^{-1}(z)$. Thus, we start by studying the semi-simple case.

From now on, we do not assume anymore that the graphs we consider on
$M$ are simple.

\subsection{The case of a semi-simple structure group}

Assume that $G$ is semi-simple and endow it with its unit-volume
bi-invariant Riemannian metric $\gamma$. Let $dg$ be the corresponding
Riemannian volume. Let also $p:\R^*_+ \times G \lra \R^*_+$ be the
fundamental solution on $G$ of the heat equation $\frac{1}{2}\Delta
-\partial_t$.

Let $\G=(\V,\E,\F)$ be a graph on $M$. Choose an orientation
$\E^+=\{e_1,\ldots,e_r\}$ of $\G$. Let $F$ be a face of this graph and
$e_{i_1}^{\epsilon_1} \ldots e_{i_n}^{\epsilon_n}$ a loop which
represents $\partial F$, where $\epsilon_1,\ldots, \epsilon_n=\pm 1$. 

Given an arbitrary group $X$, we define the mapping $h_{\partial F}^X
: X^{\E^+} \lra X/\Ad$ by setting $h_{\partial F}^X(x_1,\ldots, x_r)=
[x_{i_n}^{\epsilon_n}\ldots x_{i_1}^{\epsilon_1}]$, where $\Ad$ is the
adjoint action of $X$ on itself and $[x]$ denotes the conjugacy class of $x$.

We define also, for each path $c=e_{i_1}^{\epsilon_1} \ldots
e_{i_n}^{\epsilon_n}$ in $\E^*$, the discrete holonomy along $c$ as
the mapping $h_c : X^{\E^+} \lra X$ defined by
$h_c(x_1,\ldots,x_r)=x_{i_n}^{\epsilon_n} \ldots
x_{i_1}^{\epsilon_1}$.

We shall soon need another mapping, namely $\tilde h_{\partial F} :
\wG^{(\E)} \lra \wG/\Ad$ which is defined by setting $\tilde
h_{\partial F} (\tilde g)=[\tilde g_{e_{i_n}^{\epsilon_n}} \ldots
\tilde g_{e_{i_1}^{\epsilon_1}}]$. With this definition, $\tilde
h_{\partial F}(\tilde H_\G(s,\omega))$ is the conjugacy class of the
holonomy of $s_F^*\omega$ along $\partial F$.

Recall that the heat kernel is constant on conjugacy classes. The
usual discrete Yang-Mills measure on $G^{\E^+}$ at temperature $T>0$
is the Borel probability measure $P^\G_T$ defined by
\be \label{bas T}
dP^\G_T = \frac{1}{Z^\G_T} \prod_{F\in \F}
p_{T\sigma(F)} \circ h^G_{\partial F} \; dg^{\E^+},
\ee
where $Z^\G_T$ is the normalization constant (see for example
\cite{Levy_AMS}).

Endow $\wG$ with the Riemannian metric $\pi^*\gamma$, where $\pi:\wG
\lra G$ is the covering map. Let $d\tilde g$ be the corresponding
Riemannian volume. Observe that $\int_\wG d\tilde g = |\Pi|$. 

The configuration space $\wG^{(\E)}$ is a closed subgroup of the
compact Lie group $\wG^\E$. It is not connected unless $G$ is simply
connected, in which case the present work is pointless. In any case,
it carries bi-invariant measures with finite total mass. Let $\lambda$
denote the one such that $\lambda(\wG^{(\E)})=|\Pi^{\E}|$. Let us
identify $\wG^{(\E)}$ with a subgroup of $\wG^{2r}$, according to
$\tilde g=(\tilde g_{e_1},\tilde g_{e_1^{-1}},\ldots, \tilde g_{e_r},
\tilde g_{e_r^{-1}})$.

\begin{lemma}
Let $f$ be a continuous function on $\wG^{(\E)}$. Then
\be\label{mesure invariante}
\int_{\wG^{(\E)}} f \; d\lambda = \sum_{z_1,\ldots,z_r \in
\Pi} \int_{\wG^r} f(\tilde g_1, \tilde g_1^{-1} z_1, \ldots, \tilde g_r,
\tilde g_r^{-1} z_r) \; d\tilde g_1 \ldots d\tilde g_r.
\ee
\end{lemma}

\pf The right-hand side of the expression above defines a bi-invariant
measure with total mass $|\Pi|^{2r}=|\Pi^\E|$ on $\wG^{(\E)}$,
which then must be $\lambda$. \qed

Let $\tilde p: \R^*_+ \times \wG \lra \R^*_+$ be the fundamental
solution of the heat equation on $\wG$. It is related to the heat
kernel on $G$ as follows.

\begin{lemma} \label{noyau de la chaleur}
For all $t>0$ and all $\tilde g \in \wG$, the
following relation holds:
$$\sum_{z\in\Pi} \tilde p_t(\tilde g z) = p_t(\pi(\tilde g)).$$
\end{lemma}

\pf This follows immediately from the fact that both $G$ and $\wG$ are
endowed with their Riemannian volumes and $\pi : \wG \lra G$ is a
local isometry. \qed

Most of the following proposition consists in definitions.

\begin{proposition} Let $T>0$ be a positive number. Define the discrete
Yang-Mills measure on $\wG^{(\E)}$ at temperature $T$ as the Borel
probability measure $\tilde P^\G_T$ given by 
\be 
\label{mesure mixte} 
\frac{d\tilde P^\G_T}{d\lambda} = \frac{1}{\tilde Z^\G_T}
\prod_{F\in \F} \tilde p_{T\sigma(F)}\circ \tilde h_{\partial F}.  
\ee
Choose $T>0$ and $z \in \Pi$. One has $\tilde P^\G_T(\{\tilde g \in
\wG^{(\E)} \; | \; \oo(\tilde g) = z\}) >0$. Define the probability
$\tilde P^\G_{T,z}$ on $\wG^{(\E)}$ by 
\be
\label{mesure discrete en haut} \tilde P^\G_{T,z} = \frac{1}{\tilde
Z^\G_{T,z}} \1_{\oo^{-1}(z)} \; \tilde Z^\G_T \tilde P^\G_T.  
\ee
The following relation holds:
\be \label{dec conv}
\sum_{z\in \Pi} \tilde Z^\G_{T,z} \tilde P^\G_{T,z} =
\tilde Z^\G_T \tilde P^\G_T.
\ee
Let $z$ be an element of $\Pi\simeq \pi_1(G)$. The
discrete Yang-Mills measure on $\G$ at temperature $T>0$ and for a
bundle of type $z$ is the Borel probability measure on $G^{\E^+}$ defined by
\be \label{mesure discrete}
P^\G_{T,z} = \pi_* \tilde P^\G_{T,z},
\ee
where $\pi: \EC \lra G^{\E^+}$ is the natural projection. 
For all $T>0$ and $z\in\Pi$, the measure $\tilde P^\G_{T,z}$
(resp. $P^\G_{T,z}$) is invariant under the action of $\wJ_\G$
(resp. $\J_\G$).
\end{proposition}

\pf Let us prove that $\tilde P^\G_T(\{\tilde g \in \wG^{(\E)} \; | \;
\oo(\tilde g) = z\}) >0$. The density of $\tilde P^\G_T$ is a smooth
positive function on $\wG^{(\E)}$. It is thus enough to prove that
$\lambda(\oo^{-1}(z))$ is positive. This follows from the fact that
$\oo^{-1}(z)$ is the non-empty union of some of the finitely many
connected components of $\wG^{(\E)}$.

That the decomposition (\ref{dec conv}) is true is a
straightforward consequence of the definitions. Observe in fact that
it is a decomposition into mutually singular parts. 

There remains to prove the claimed gauge-invariance of the
measures. Choose $T>0$ and $z\in\Pi$. For each face $F$ of the graph,
one checks easily that the action of $\wJ_\G$ on $\EC$ leaves the
mapping $\tilde h_{\partial F} :\EC \lra \wG/\Ad$ invariant. According
to Lemma \ref{invariance obstruction}, it preserves also the mapping
$\oo : \EC \lra \Pi$. Hence, the measures $\tilde P^\G_T$ and $\tilde
P^\G_{T,z}$ are left invariant by $\wJ_\G$. One deduces the
corresponding assertions for $P^\G_T$ and $P^\G_{T,z}$ by using the
equivariance properties of $\pi:\EC \lra G^{\E^+}$. \qed

We shall now give an expression of the measures $P^\G_{T,z}$ at the
level of $G^{\E^+}$.

\begin{proposition} \label{prop: mesure discrete}
Let $g=(g_1,\ldots,g_r)$ be an element of
$G^{\E^+}$. Let $z$ be an element of $\Pi$. Let $\tilde g=(\tilde
g_1,\ldots,\tilde g_r)$ be a lift of $g$ to $\wG^{\E^+}$. Then the
number
\be\label{defD}
\sum_{\stackrel{\scriptstyle (z_F)_{F}\in \Pi^\F}{\prod z_F=z}} \prod_{F\in \F} \tilde
p_{T \sigma (F)} (h^{\wG}_{\partial F}(\tilde g) z_F)
\ee
does not depend on the choice of $\tilde g$. We denote it by
$D^\G_{T,z}(g)$.

Moreover, the measure $P^\G_{T,z}$ satisfies
\be \label{expression mesure discrete}
dP^\G_{T,z} = \frac{1}{Z^\G_{T,z}} D^\G_{T,z} \; dg^{\otimes \E^+}.
\ee

Finally, one has $\tilde Z^\G_{T,z}=|\Pi|^{|\E|-|\F|}  Z^\G_{T,z}$.
\end{proposition}

\pf Let $w$ be an element of $\Pi$. Let us look at the effect of
replacing for example $\tilde g_1$ by $\tilde g_1 w$ in
(\ref{defD}). Set $F_1=L(e_1)$ and $F_2=L(e_1^{-1})$. Then this is
equivalent to replacing $(z_{F_1},z_{F_2})$ by $(z_{F_1}w,
z_{F_2}w^{-1})$ and it does not change the value of the sum. This
proves the first assertion.

To prove the second one, consider a continuous function $f$ on
$G^{\E^+}$. Then, by definition of $P^\G_{T,z}$, 
\be\label{c1}
\int_{G^{\E^+}} f(g) \; dP^\G_{T,z} = \frac{1}{\tilde Z^\G_T}
\sum_{\stackrel{z_1,\ldots,z_r \in \Pi}{\scriptscriptstyle z_1\ldots z_r
=z}} \int_{\wG^r} f(\pi (\tilde g)) \prod_{F\in \F} \tilde
p_{T\tilde \sigma(F)}\circ \tilde h_{\partial F} (\tilde
g_1,\tilde g_1^{-1} z_1,\ldots,\tilde g_r,\tilde g_r^{-1} z_r)\;
d\tilde g.
\ee
Now observe that, for each face $F$, 
$$\tilde h_{\partial F} (\tilde g_1,\tilde g_1^{-1} z_1,\ldots,\tilde
g_r,\tilde g_r^{-1} z_r) =h^{\wG}_{\partial F}(\tilde
g_1,\ldots,\tilde g_r) \prod_{i\in [1,r] : L(e_i^{-1})=F} z_i.$$ 
Set $z_F=\prod_{i \in [1,r] : L(e_i^{-1})=F} z_i$. For two distinct
faces $F$ and $F'$, $z_F$ and $z_{F'}$ are products of disjoint
collections of $z_i$'s. Actually, the sets $\{ i \in [1,r] :
L(e_i^{-1})=F \}$ form a partition of $[1,r]$. Since the image of the
uniform measure on the product of a finite number of copies of $\Pi$ by
multiplication of the factors is the uniform measure on $\Pi$, the
right hand side of (\ref{c1}) equals
\begin{eqnarray*}
&& \frac{|\Pi^{\E^+}|}{\tilde Z^\G_T |\Pi^{\F}|} \int_{\wG^r} f(\pi(\tilde g))
\sum_{\stackrel{(z_F)\in \Pi^{\F}}{\prod z_F = z}} \prod_{F\in \F} 
\tilde p_{T\tilde \sigma(F)}(h^{\wG}_{\partial F} (\tilde
g_1,\ldots,\tilde g_r) z_F) \; d\tilde g\\
&=& \frac{|\Pi|^{|\E|-|\F|}}{\tilde Z^\G_T} \int_{G^r} f(g) D^\G_{T,z}(g)
\; dg.
\end{eqnarray*}
The second assertion is proved, as well as the third. \qed

\begin{remark} Let $f$ be a continuous function on $G^{\E^+}$. The
integral of $f$ with respect to $P^\G_{T,z}$ can be written as
follows. 
\begin{eqnarray*}
\int_{G^{\E^+}} f \; dP^\G_{T,z} &=& \frac{1}{Z^\G_{T,z}}
\int_{G^{\E^+}} f(g) \sum_{\stackrel{\scriptstyle (z_F)_{F}\in
\Pi^\F}{\prod z_F=z}} \prod_{F\in \F} \tilde p_{T \sigma (F)}
(h^{\wG}_{\partial F}(\tilde g) z_F) \; dg  \\
&=& \frac{1}{Z^\G_{T,z}} \frac{1}{|\Pi^{\E^+}|} \int_{\wG^{\E^+}}
f(\pi(\tilde g)) \sum_{\stackrel{\scriptstyle (z_F)_{F}\in
\Pi^\F}{\prod z_F=z}} \prod_{F\in \F} \tilde p_{T \sigma (F)}
(h^{\wG}_{\partial F}(\tilde g) z_F) \; d\tilde g \\ 
&=& \frac{1}{Z^\G_{T,z}} \frac{1}{|\Pi^{\E^+}|}
\sum_{\stackrel{\scriptstyle (z_F)_{F}\in \Pi^\F}{\prod z_F=z}}
\int_{\wG^{\E^+}} f(\pi(\tilde g))\prod_{F\in \F} \tilde p_{T \sigma (F)}
(h^{\wG}_{\partial F}(\tilde g) z_F) \; d\tilde g.
\end{eqnarray*}
Thanks to the invariance by translation of the Haar measure, all the
terms in the last sum are equal. This statement is also the content of
\cite[Lemma 7.5]{Sengupta_AMS}. Hence, in particular, if we choose a
face $F_*$ in $\F$, then
$$\int_{G^{\E^+}} f \; dP^\G_{T,z} =
\frac{1}{Z^\G_{T,z}}\frac{|\Pi|^{|\F|-1}}{|\Pi^{\E^+}|}
\int_{\wG^{\E^+}} f(\pi(\tilde g)) \tilde
p_{T\sigma(F_*)}(h^{\wG}_{\partial F_*}(\tilde g)z) \prod_{F\in
\F-\{F_*\}} \tilde p_{T \sigma (F)} (h^{\wG}_{\partial F}(\tilde g)
z_F) \; d\tilde g.$$ 
Comparing this expression with \cite[Theorem 8.4]{Sengupta_AMS} shows
that our definition of the discrete Yang-Mills measure associated to a
specific isomorphism class of $G$-bundles is consistent with that
previously given by A. Sengupta.
\end{remark}

\begin{corollary} The covering map $\pi : \wG^{(\E)} \lra G^{\E^+}$
satisfies $\pi_* \tilde P^\G_T = P^\G_T$. Moreover, $\tilde Z^\G_T =
|\Pi|^{|\E|-|\F|} Z^\G_T$ and $\displaystyle \sum_{z\in \Pi} Z^\G_{T,z}
P^\G_{T,z} = Z^\G_T P^\G_T$.
\end{corollary}

\pf By definition of $D^\G_{T,z}$ and by Lemma \ref{noyau de la
chaleur},
$$\sum_{z\in \Pi} D^\G_{T,z}(g) = \prod_{F\in \F} p_{T
\sigma(F)}(h^G_{\partial F}(g)).$$
Hence, by Proposition \ref{prop: mesure discrete}, $ \sum_{z\in \Pi}
Z^\G_{T,z} P^\G_{T,z} = Z^\G_T P^\G_T.$ On the other hand, 
$$\sum_{z\in \Pi} Z^\G_{T,z} P^\G_{T,z} = \frac{1}{|\Pi|^{|\E|-|\F|}}\; \pi_*
\left[ \sum_{z\in \Pi} \tilde Z^\G_{T,z} \tilde P^\G_{T,z} \right] =
\frac{1}{|\Pi|^{|\E|-|\F|}}\; \pi_* (\tilde Z^\G_T \tilde P^\G_T).$$
The result follows. \qed

\subsection{The general case}
\label{Discret cas general}

Let us drop the assumption that $G$ is semi-simple. Then, except in
trivial cases like $\E=\varnothing$, definitions (\ref{mesure
mixte}) and (\ref{mesure discrete en haut}) do not make sense anymore,
because they involve infinite normalization constants. Fortunately,
Proposition \ref{prop: mesure discrete} is still meaningful, as the
following result shows.

\begin{lemma} Let $T>0$ and $z\in \Pi$. The definition of $D^\G_{T,z}$
(see Proposition \ref{prop: mesure discrete}) makes sense on any
compact connected Lie group. If $G$ is such a group, the function thus
defined is bounded on $G$.
\end{lemma}

%It is possible to work on a slice of $\wG^{(\E)}$ under the action of
%$\wJ_\G$ and perform an analysis very similar to what we have done in
%the previous section. However, this requires unpleasant choices and
%does not seem to be very illuminating. We prefer to prove directly
%that Proposition \ref{prop: mesure discrete} makes sense in general
%and take it as a definition. We shall then prove that, in the Abelian
%case, this is equivalent to the bundle-dependent definition suggested
%in \cite{L}.

\pf The trouble is that $D^\G_{T,z}$ might take infinite
values. However, for all $g \in G^{\E^+}$, and by Lemma \ref{noyau de
la chaleur}, the sum of positive functions $\sum_{z \in \Pi}
D^\G_{T,z}$ is equal to $\prod_{F \in \F} p_{T\sigma(F)} \circ
h^G_{\partial F}$, which is the density of the usual discrete
Yang-Mills measure and is finite and even bounded on $G$. The result
follows. \qed

\begin{definition} Let $\G$ be a graph on $M$. Let $z$ be an element
of $\Pi$. Let $T$ be a positive real number. Then the discrete
Yang-Mills measure on $\G$ associated with the isomorphism class of
bundles corresponding to $z$ and at temperature $T$ is the Borel
probability measure $P^\G_{T,z}$ on $G^{\E^+}$ defined by
$$dP^\G_{T,z} = \frac{1}{Z^\G_{T,z}} D^\G_{T,z} \; dg^{\otimes
\E^+}.$$
\end{definition}

The following lemma follows immediately from Lemma \ref{noyau de la chaleur}.

\begin{lemma}\label{convexe}
The relation $\displaystyle \sum_{z\in \Pi} Z^\G_{T,z}
P^\G_{T,z} = Z^\G_T P^\G_T$ holds.
\end{lemma}

At this point, we have written the usual discrete Yang-Mills measure
as a convex combination of probability measures, one for each
isomorphism class of principal $G$-bundle over $M$. In the case of a
semi-simple structure group and a simple graph, this decomposition has
been given a strong geometrical motivation. In the next subsection, we
check that our construction is consistent with some observations made
in \cite[Chapter 3]{Levy_AMS} in the case of an Abelian structure
group.

\subsection{The case of an Abelian structure group}

Let us compare our definition of the discrete Yang-Mills measures with
the study of the Abelian case presented in \cite[Section 1.9]{Levy_AMS}. As
explained in \cite[Lemma 1.34]{Levy_AMS}, when $G$ is Abelian, the
non-trivial information about any gauge-invariant measure on
$G^{\E^+}$ is contained in the law of the discrete holonomies along
the boundaries of the faces. This is why, in what follows, we focus on
this law.

Assume that $G=SO(2)^m$. Let $e: \R^m \lra SO(2)^m$ be defined by
$e(x_1,\ldots,x_m)=(e^{2i\pi x_1},\ldots,e^{2i\pi x_m})$. Assume that
$\F=\{F_1,\ldots,F_n\}$. For each $i=1 \ldots n$, set
$\sigma_i=\sigma(F_i)$. Set also $\sigma_M = \sigma(M)$. Let
$Y_1,\ldots,Y_n$ be independent $\R^m$-valued Gaussian random
variables with $Y_i \sim \N(0,T\sigma_i I_m)$ for each $i$, where
$I_m$ is the identity matrix. Let $S=Y_1+ \ldots +Y_n$ be their
sum. For each $i=1\ldots n$, set $X_i=Y_i- \frac{\sigma_i}{\sigma_M}
S$.

Let $Z$ be a $\Z^m$-valued random variable, independent of
$Y_1,\ldots,Y_n$, such that, for all $z \in \Z^m$,
$$\P(Z=z)= C \; \exp {-\frac{|z_1|^2+\ldots+|z_m|^2}{2T\sigma_M}},$$ where
$C$ is the correct normalization constant.

\begin{theorem} For all $T>0$, all $z\in \Z^m$, all $f$ continuous on
$G^{\E^+}$, one has
\be \label{identification abelienne}
\int_{G^{\E^+}} f(h_{\partial {F_1}},\ldots,h_{\partial {F_n}}) \;
dP^\G_{T,z} = \E\left[ f\left(e(X_1+ \frac{\sigma_1}{\sigma_M}
z),\ldots,e(X_n+\frac{\sigma_1}{\sigma_M} z)\right) \right].
\ee
Moreover,
\be \label{abelien global}
\int_{G^{\E^+}}
f(h_{\partial {F_1}},\ldots,h_{\partial {F_n}}) \; dP^\G_T = \E\left[
f\left(e(X_1+ \frac{\sigma_1}{\sigma_M}
Z),\ldots,e(X_n+\frac{\sigma_1}{\sigma_M} Z)\right) \right].  \ee
\end{theorem}

The second relation is proved in the case $T=m=1$ in \cite[Proposition
1.38]{Levy_AMS}. We observed there that $Z$ plays the role of a total
curvature, that is, of the obstruction class of the bundle, and that
replacing $Z$ by a deterministic element of $\Z^m$ would be equivalent
to selecting an isomorphism class of $SO(2)^m$-bundles. We prove now that
our definition of $P^\G_{T,z}$ is consistent with this observation. \\

\pf The proof is very similar to that of \cite[Prop. 1.38]{Levy_AMS}. For the
convenience of the reader and because \cite{Levy_AMS} deals with
a more restrictive situation, we present a detailed sketch of proof.

By using the definition of $P^\G_{T,z}$ and the fact that, under the
Haar measure on $G^{\E^+}$, the holonomies along the boundaries of all
faces except one are independent and uniformly distributed on $G$, one
finds that the left hand side of (\ref{identification abelienne}) is
equal to
\be 
\left(
(2\pi)^{n-1} \frac{\sigma_1 \ldots \sigma_n}{\sigma_M}
\right)^{-\frac{m}{2}} e^{\frac{|z|^2}{2T\sigma_M}}
\int_{([0,1]^m)^{n-1}} f(e(x_1),\ldots,e(x_n)) D_L(x_1,\ldots,x_{n-1})
\; dx_1 \ldots dx_{n-1}, \ee with $x_n = -x_1 - \ldots -x_{n-1}$ and
\be D_L(x_1,\ldots,x_{n-1})= \sum_{z_1,\ldots,z_{n-1} \in \Z^m} \exp
-\frac{1}{2T} \left[\sum_{i=1}^n \frac{|x_i+ z_i|^2}{\sigma_i}\right],
\ee 
where we have set $z_n = z - z_1 - \ldots - z_{n-1}$.

By computing the covariance of the Gaussian vector $(X_1,\ldots,X_n)$,
one finds that the right hand side of (\ref{identification abelienne})
is equal to
\be
\left( (2\pi)^{n-1} \frac{\sigma_1 \ldots \sigma_n}{\sigma_M}
\right)^{-\frac{m}{2}} \int_{([0,1]^m)^{n-1}} f(e(x_1),\ldots,e(x_n))
D_R(x_1,\ldots,x_{n-1}) 
\; dx_1 \ldots dx_{n-1},
\ee 
with again $x_n = -x_1 - \ldots -x_{n-1}$ and 
\be
D_R(x_1,\ldots,x_{n-1})=\sum_{w_1,\ldots,w_{n-1} \in \Z^m} \exp
-\frac{1}{2T} \left[ \sum_{i=1}^{n-1} \frac{1}{\sigma_i} \left|
x_i+w_i - \frac{\sigma_i}{\sigma_M} \right|^2 + \frac{1}{\sigma_n}
\left| \sum_{i=1}^n x_i + w_i - \frac{\sigma_i}{\sigma_M} z \right|^2
\right],
\ee
where we have set $w_n= z - w_1 - \ldots - w_{n-1}$. 

An elementary computation shows that $\exp({\frac{|z|^2}{2T\sigma_M}})
D_L= D_R$ and this is most easily seen by identifying $w_i$ and $z_i$
for $i=1\ldots n-1$. 

This proves (\ref{identification abelienne}) and (\ref{abelien
global}) follows by summing both sides over $z\in \Z^m$, with a weight
$\P(Z=z)$.  \qed

%Notice that integrals over $[0,1]$ have appeared in the proof. This is
%an example of what was alluded to earlier as ``working on a slice of the
%configuration space under the action of the gauge group'', made
%necessary by the non-compactness of $\R^m$.

\section{The continuous measures}

%We recall some notation from the beginning of Section \ref{Discret cas
%general}. Let $G$ be a compact connected Lie group, $\pi: \wG \lra G$
%a universal cover of $G$, $T>0$ a real number and $z$ an element of
%$\Pi = \ker \pi$.

\subsection{Invariance under subdivision} 

Let $\G_1=(\V_1,\E_1,\F_1)$ and $\G_2=(\V_2,\E_2,\F_2)$ be two graphs
on $M$, not necessarily simple ones. We say that $\G_2$ is finer than
$\G_1$ if $\E_1^* \subset \E_2^*$, or equivalently $\E_1 \subset
\E_2^*$. We denote this by $\G_1 \leq \G_2$.

Let $\E_1^+$ and $\E_2^+$ be orientations of $\G_1$ and $\G_2$
respectively. We have two
probability spaces $(G^{\E_1^+}, P^{\G_1}_{T,z})$ and
$(G^{\E_2^+},P^{\G_2}_{T,z})$, and there is a natural mapping $f_{\G_1\G_2}$
from the second one to the first one, defined by
$f_{\G_1\G_2}(g)=(h_e(g))_{e\in \E_1}$.

\begin{proposition}\label{inv sub}
 The mapping $f_{\G_1\G_2} : G^{\E_2^+} \lra G^{\E_1^+}$
is onto. Moreover, for all $T>0$ and $z\in\Pi$, it satisfies 
\be \label{projectivite}
(f_{\G_1\G_2})_* P^{\G_2}_{T,z} = P^{\G_1}_{T,z}.
\ee
\end{proposition}

\pf That $f_{\G_1\G_2}$ is onto has been proved in \cite[Theorem
1.22]{Levy_AMS} and is anyway easy to check.

The proof of (\ref{projectivite}) is also essentially the same than
that of \cite[Theorem 1.22]{Levy_AMS}. We give a detailed sketch of proof and
give full details for the only non-trivial and new step.  First, one
proves that there exists a finite sequence $\G_1=\G'_1 \leq \ldots
\leq \G'_n = \G_2$ of graphs such that, for each $k=1\ldots n-1$, one
can deduce $\G'_{k+1}$ from $\G'_k$ by one of the following elementary
operations: 
\begin{itemize}
\item $V$~:  Adding a vertex in the middle of an edge. 
\item $E_1$~: Adding a `loose' edge, that is, an edge such that exactly one
of its two endpoints already belong to the graph.
\item $E_2$~: Adding a `tight' edge, that is, joining two vertices by a
new edge.
\end{itemize}

There is a probability space associated to each of them and, for each
$k$, a mapping $f_{\G'_k\G'_{k+1}}$ from the space associated to
$\G'_{k+1}$ to the one associated to $\G'_k$. In fact, $f_{\G_1\G_2}=
f_{\G'_1 \G'_2} \circ \ldots \circ f_{\G'_{n-1} \G'_n}$ and it is
sufficient to prove the result when $\G_2$ can be deduced from $\G_1$
by one of the elementary operations described above.

In the cases of operations $V$ and $E_1$, the result follows
immediately from the basic properties of the Haar measure. The proof
in the case of operation $E_2$ involves the properties of the heat
kernel.

Assume that $\E_1^+=\{e_1,\ldots,e_{r-1}\}$ and
$\E_2^+=\{e_1,\ldots,e_r\}$, with $\{\underline{e_r}, \overline{e_r}\}
\subset \V_1$. We claim that, for all $g_1,\ldots,g_{r-1} \in G$, 
\be
\label{case E2} \int_{G} D^{\G_2}_{T,z}(g_1,\ldots,g_r) \; dg_r =
 D^{\G_1}_{T,z}(g_1,\ldots,g_{r-1}).  
\ee 
The edge $e_r$ cuts a face of $\F_1$ into two faces. Set
$\F_1=\{F_1,\ldots,F_n,F\}$ and $\F_2=\{F_1,\ldots,F_n,F',F''\}$, with
$F'\cap F''=e_r$ and $F'\cup F''=F$. Assume that $L(e_r)=F'$ and
$L(e_r^{-1})=F''$. Then there exists two paths $c'$ and $c''$ in
$\E_1^*$ such that $e_r c'$ is the boundary of $F'$, $c'' e_r^{-1}$
that of $F''$ and $c'c''$ that of $F$. Let us fix $\tilde g=(\tilde
g_1,\ldots,\tilde g_{r-1})$ some arbitrary lift of
$(g_1,\ldots,g_{r-1})$. Then the left hand side of (\ref{case E2}) is
equal to
\begin{eqnarray*}
&&\hskip -1cm \sum_{z_{1},\ldots,z_{n}\in \Pi}
\prod_{i=1}^n \tilde p_{T\sigma(F_i)}(h^{\wG}_{\partial
F_i}(\tilde g)z_i)  \int_{G} \sum_{\pi(\tilde x)=x}\tilde
p_{T\sigma(F')}(h^{\wG}_{c'} (\tilde g) \tilde x) \tilde
p_{T\sigma(F'')} (\tilde x^{-1} h^{\wG}_{c''} (\tilde g)
z \prod_{i=1}^n z_i^{-1} ) \; dx \\
&=& \sum_{z_{1},\ldots,z_{n}\in \Pi}
\prod_{i=1}^n \tilde p_{T\sigma(F_i)}(h^{\wG}_{\partial
F_i}(\tilde g)z_i) \int_\wG \tilde
p_{T\sigma(F')}(h^{\wG}_{c'} (\tilde g) \tilde x) \tilde
p_{T\sigma(F'')} (\tilde x^{-1} h^{\wG}_{c''} (\tilde g)
z \prod_{i=1}^n z_i^{-1} ) \; d\tilde x \\
&=& \sum_{z_{1},\ldots,z_{n}\in \Pi}
\prod_{i=1}^n \tilde p_{T\sigma(F_i)}(h^{\wG}_{\partial
F_i}(\tilde g)z_i)\; \tilde p_{T
\sigma(F)}(h^{\wG}_{c'c''}(\tilde g)z \prod_{i=1}^n z_i^{-1})\\
&=& D^{\G_1}_{T,z}(g_1,\ldots,g_{r-1}).
\end{eqnarray*}
We have used the convolution property of the heat kernel between the
 second and the third line. This finishes the proof in the case of a
 transformation $E_2$. \qed

\subsection{Random holonomy along piecewise geodesic paths}

Recall that $M$ is endowed with a Riemannian metric. Let $\Pi M$
denote the set of piecewise geodesic paths on $M$, so that $\Pi M
\subset PM$. Let ${\cal G}$ denote the set of graphs with geodesic
edges.

Recall \cite[Section 2.3]{Levy_AMS} that $({\cal G},\leq)$ is a directed set,
that is, a partially ordered set such that any two elements admit an
upper bound. This is a consequence of the rigidity of geodesics and it
is the property which makes ${\cal G}$ an interesting set for
us. Recall that, if $J$ is a subset of $PM$ stable by concatenation,
$f : J \lra G$ is said to be multiplicative if, whenever $c_1$ and
$c_2$ belong to $J$ and satisfy $\overline{c_1}=\underline{c_2}$, one
has $f(c_1c_2)=f(c_2)f(c_1)$.

The projective limit of the system $((G^{\E^+})_{\G \in {\cal G}},
(f_{\G\G'})_{\G\leq \G'})$ is canonically isomorphic to the set
$\M(\Pi M,G)$ of multiplicative functions from $\Pi M$ to $G$, with
projection mappings $f_\G : \M(\Pi M,G) \lra G^{\E^+}$ given by
restriction. 

According to Proposition \ref{inv sub} and general results on
projective limits of measure spaces (see \cite{Rao}), it is possible
to take the projective limit of the compact Borel probability spaces
$(G^{\E^+},P^\G_{T,z})_{\G\in {\cal G}}$ with respect to the mappings
$(f_{\G\G'})_{\G\leq \G'}$.

\begin{proposition} Let $\C$ be the cylinder $\sigma$-field of $\M(\Pi
M,G)$. There exists on the measurable space $(\M(\Pi M,G),\C)$ a
unique probability measure $P_{T,z}$ such that, for all graph $\G$
with geodesic edges, $(f_\G)_* P_{T,z}=P^\G_{T,z}$.
\end{proposition}

Let $(H_\zeta)_{\zeta \in \Pi M}$ denote the canonical process on
$(\M(\Pi M,G),\C)$. Let us also denote, for each piecewise geodesic
path $\zeta$, by $\ell(\zeta)$ the length of $\zeta$.  Finally, let
$d_G$ denote the Riemannian distance on $G$. The main property of
$P_{T,z}$ is the following.

\begin{proposition} \label{small loops}
There exist two constants $K,L>0$ such that, for
any loop $\zeta \in \Pi M$,  $\ell(\zeta) \leq L$ implies $\E_{P_{T,z}}
[d_G(1,H_\zeta)] \leq K \ell(\zeta)$.
\end{proposition}

\pf Assume first that $\zeta$ is a simple loop. Then, if $L$ is small
enough, $\ell(\zeta) \leq L$ implies that $\zeta$ is homotopic to a
constant loop and, by a local isoperimetric inequality
\cite[Proposition 2.15]{Levy_AMS}, bounds a domain $D \subset M$ such that
$\sigma(D) \leq K_1 \ell(\zeta)^2$, where $K_1$ depends only on $L$.
Since the Riemannian metric is smooth, $K_1$ remains bounded when $L$
gets smaller and we may assume, by taking $L$ small enough, that $K_1
L^2 \leq \frac{1}{2} \sigma(M)$.

The law of $H_\zeta$ can be computed in any graph $\G$ such that
$\zeta$ belongs to $\E^*$. We choose a graph with only two faces,
namely $\overline D$ and $M\backslash D$. Then, if $g$ denotes the
genus of $M$, the expectation we want to estimate is equal to
$$\frac{1}{Z^\G_{T,z}}\int_{G^{2g+1}} d_G(1,x) \;
\sum_{\pi(\tilde x)=x} \tilde p_{T \sigma(D)}(\tilde x) \tilde
p_{T \sigma(D^c)}(\tilde x^{-1} [a_1,b_1]\tilde{\;} \ldots
[a_g,b_g]\tilde{\;} z) \; dx da_1 db_1 \ldots da_g db_g,$$ 
where $[a,b]\tilde{\;}$ denotes the lift to $\wG$ of $[a,b]$, that is, the
value of $[\tilde a,\tilde b]$ for any lift $(\tilde a,\tilde b)$ of
$(a,b)$. By the convolution property of the heat kernel,
$$Z^\G_{T,z}= \int_{G^{2g}} \tilde p_{T 
\sigma(M)}([a_1,b_1]\tilde{\;} \ldots [a_g,b_g]\tilde{\;} z) \; da_1 db_1
\ldots da_g db_g,$$
which does not depend on $\G$ nor on $\zeta$. Hence,

\begin{eqnarray*}
\E_{P_{T,z}} [d_G(1,H_\zeta)] &\leq & K_2 \parallel \tilde
p_{\frac{T}{2} \sigma(M)} \parallel_{\infty} \int_G
d_G(1,x) \; \sum_{\pi(\tilde x)=x} \tilde p_{T\sigma(D)}(\tilde
x) \; dx \\ 
&\leq& K_3 \int_G d_G(1,x) p_{T\sigma(D)}(x) \; dx \\ 
&\leq & K_4 \sqrt{\sigma(D)} \\ 
&\leq & K \ell(\zeta).
\end{eqnarray*}

We have used Lemma \ref{noyau de la chaleur} to pass from $\tilde p$
to $p$ and then a classical estimation on the heat kernel, see for
example \cite[Proposition 1.31]{Levy_AMS}.

The case where $\zeta$ is not assumed to be a simple loop anymore
follows now easily along the lines of \cite[Sections 2.4 and
2.5]{Levy_AMS}. \qed

\subsection{The Yang-Mills measures}

Once Proposition \ref{small loops} is proved, we can go through the
construction of the Yang-Mills measure as in \cite[Sections 2.6,2.7
and 2.10]{Levy_AMS}. For the convenience of the reader, we recall the main
steps and indicate where the proofs can be found.

The topology on $PM$ is that of convergence in length, that is,
uniform convergence plus convergence of the length. It is induced by
the distance $d_\ell(c,c')=\inf \sup_{t\in [0,1]} d(c(t),c'(t)) + |
\ell(c) - \ell(c')|$, where the infimum is taken over all
parametrizations of $c$ and $c'$.\\

1. One defines for each $c \in PM$ a random variable $H_c$
   characterized by the fact that, if $(\zeta_n)_{n\geq 0}$ is a
   sequence of $\Pi M$ converging in length to $c$ and such that each
   $\zeta_n$ shares the same endpoints as $c$, then
   $\E_{P_{T,z}}[d_G(H_{\zeta_n},H_c)] \to 0$. \cite[Proposition
   2.35, Section 2.6.4]{Levy_AMS}

2. One checks that the same convergence property holds without the
   assumption that the approximating paths are piecewise
   geodesic. \cite[Proposition 2.42]{Levy_AMS}

3. One checks that the finite-dimensional marginals of the family
   $(H_c)_{c\in PM}$ are consistent with the discrete theory. More
   precisely, for each graph $\G$, not necessarily piecewise geodesic,
   the law of $(H_e)_{e\in\E^+}$ is $P^\G_{T,z}$. As a side result one
   gets the fact that the number $Z^\G_{T,z}$ is independent of
   $\G$. \cite[Proposition 2.46,2.50]{Levy_AMS}

4. Then, one considers, for each finite subset $I$ of $PM$, the
   probability spaces $(\M(I,G),P^I_{T,z})$, where $P^I_{T,z}$ is the
   law of $(H_c)_{c\in I}$ together with the natural projections
   $\M(I,G) \lra \M(J,G)$ defined whenever $J\subset I$. The
   projective limit of these probability spaces is canonically
   isomorphic to $(\M(PM,G),\C)$, where $\C$ is the cylinder
   $\sigma$-field, endowed with a probability measure that we denote
   by $P_{T,z}$. \cite[Theorem 2.62]{Levy_AMS}

\begin{theorem} Let $M$ be a compact surface without boundary endowed
   with a volume $2$-form $\sigma$. Let $G$ be a compact connected Lie
   group and $\pi:\wG \lra G$ a universal cover of $G$. Let $P \lra M$
   be a principal $G$-bundle over $M$ with obstruction class $z\in
   \ker \pi$. Let $T>0$ be a positive real number. Let $(\M(PM,G),\C)$
   denote the set of multiplicative functions from $PM$ to $G$ endowed
   with the cylinder $\sigma$-field. Let $(H_c)_{c\in PM}$ denote the
   evaluation process on this space.

There exists on $(\M(PM,G),\C)$ a unique probability measure $P_{T,z}$
such that the following two properties hold.

1. For all graph $\G=(\V,\E,\F)$ on $M$, with a choice of orientation
   $\E^+=\{e_1,\ldots,e_r\}$, the law of $(H_{e_1},\ldots,H_{e_r})$
   under $P_{T,z}$ is equal to $P^\G_{T,z}$.

2. Whenever $c$ belongs to $PM$ and $(c_n)_{n\geq 0}$ is a sequence of
   $PM$ converging in length to $c$ such that for each $n\geq 0$,
   $c_n$ and $c$ share the same endpoints,
   $\E_{P_{T,z}}[d_G(H_{c_n},H_c)]$ tends to $0$ as $n$ tends to
   $\infty$.
\end{theorem}

Recall from Lemma \ref{convexe} that, for each graph $\G$ and each
$T>0$, one has $\sum_{z\in \Pi} Z^\G_{T,z} P^\G_{T,z} = Z^\G_T
P^\G_T$. As stated above, the numbers $Z^\G_{T,z},Z^\G_T$ do not
depend on the graph $\G$. Hence, by performing simultaneously the
construction of the measure $P_{T,z}$ for each $z\in \Pi$ and also of
the measure $P_T$ as defined in \cite{Levy_AMS}, on gets the following
result.

\begin{proposition} Let us keep the notation of the theorem above. Let
$P_T$ be the probability measure on $(\M(PM,G),\C)$ constructed in
\cite[Theorem 2.62]{Levy_AMS}. The following equality holds:
\be
\label{vrai convexe} \sum_{z\in \Pi} Z_{T,z} P_{T,z} = Z_T P_T.  \ee
\end{proposition}

\subsection{Mutual singularity of the measures $P_{T,z}$}

The reader may wonder why we have not taken the projective limit of
the covering probability spaces $(\EC,\tilde P^\G_{T,z})$. It seems
indeed that something has been lost by projecting the mutually
singular measures $(\tilde P^\G_{T,z})_{z\in \Pi}$ to define the
mutually absolutely continuous measures $(P^\G_{T,z})_{z\in \Pi}$.

There are at least two answers to this question. The first is, there
is no natural projective structure of the probability spaces
$(\EC,\tilde P^\G_{T,z})$, as the reader will convince himself
easily by looking at simple examples. The fact that the covering
$\EC/\wJ_\G \lra G^{\E^+}/\J_G$ has degree $|\Pi^\F|$, which even when
$G$ is semi-simple tends to infinity as $\G$ gets finer, may be an
indication of this lack of projective structure. This means that there
is no covering of the measurable space $(\M(PM,G),\C)$ which plays a
role similar to that of the spaces $\EC$.

The second answer is that such a covering would be useless, as the
following result shows.

\begin{theorem} \label{singularite}
Let $T,T'$ be two positive real numbers. Let $z,z'$ be two elements of
$\Pi$. Then the probability measures $P_{T,z}$ and $P_{T',z'}$ are
mutually singular on $(\M(PM,G),\C)$, unless $(T,z)=(T',z')$. 
\end{theorem}

The dependence in $T$ is accessory with respect to our previous
discussion but we include it here because it does not make the proof
significantly harder. A closely related question has been discussed by
Fleischhack \cite{Fleischhack_B}. The main point is thus that there
are disjoint sectors on the space $(\M(PM,G),\C)$ corresponding to
different temperatures and different isomorphism classes of bundles.
This result should be compared to the fact that, if $(W_t)_{t\geq 0}$
is a standard real Brownian motion, then, given two real numbers
$T,T'>0$, the laws of $(W_{Tt})_{t\geq 0}$ and $(W_{T't})_{t\geq 0}$
are mutually singular unless $T=T'$.

To prove Theorem \ref{singularite}, we construct a pair $(\tau,\oo)$ of
random variables on $(\M(PM,G),\C)$ with values in $[0,+\infty] \times
\Pi$ and we show that, $P_{T,z}$ almost-surely, $(\tau,\oo)=(T,z)$. 

For this, we focus on the random holonomy along a simple family of
loops which we start by defining.

Let $g$ denote the genus of $M$. Let $D$ denote the closed unit disk
in $\R^2$. Let $q:D \lra M$ denote a continuous onto mapping such that
the restriction of $q$ to the interior of $D$ is an
orientation-preserving diffeomorphism and $q$ maps the boundary of $D$
onto $2g$ loops $a_1,b_1\ldots,a_g,b_g$ which generate $\pi_1(M)$ and
such that $q(\partial D)$ is the cycle
$[b_g^{-1},a_g^{-1}]\ldots[b_1^{-1},a_1^{-1}]$, where
$[a,b]=aba^{-1}b^{-1}$.

For each $s \in [0,1]$, let $c_s$ denote the loop in $\R^2$ based at
$(1,0)$ going once counterclockwise along the circle of center
$(1-s,0)$ and radius $s$. We project the loops $c_s$ on $M$ and
re-index them. For this, let $D_s$ denote the disk bounded
by $c_s$. Set, for each $s\in [0,1]$, $A(s)=\sigma(q(D_s))$. This
defines an increasing diffeomorphism $A:[0,1] \lra [0,\sigma(M)]$ and
we set, for each $t\in [0,\sigma(M)]$, $l_t= q(c_{A^{-1}(t)})$. It is
an element of $PM$ which bounds with positive orientation the domain
$q(D_{A^{-1}(t)})$ whose area is $t$. Finally, for each $t \in
[0,\sigma(M)]$, set $X_t=H_{l_t}$.

Recall that, if $x,y \in G$, then the commutator $[\tilde x,\tilde y]$
does not depend on the choice of $\tilde x$ and $\tilde y$ such that
$\pi(\tilde x)=x$ and $\pi(\tilde y)=y$. We denote this commutator by
$[\widetilde{x,y}]$. 

%For the sake of clarity, let us introduce the
%notation $W(a_1,b_1,\ldots,a_g,b_g)=[b_g^{-1},a_g^{-1}]\tilde{\;}
%\ldots [b_1^{-1},a_1^{-1}]\tilde{\;}$, so that
%$W(H_{a_1},H_{b_1},\ldots,H_{a_g},H_{b_g})$ is a lift of the holonomy
%along the loop $[a_1,b_1]\ldots [a_g,b_g]$.  Set $\tilde T=(\deg
%\rho)^{-\frac{2}{d}} T$.

\begin{lemma} \label{bridge}
The conditional law of the family of random variables $(X_t)_{t \in
[0,\sigma(M)]}$ under $P_{T,z}$ given $(H_{a_1},H_{b_1}, \ldots,
H_{a_g},H_{b_g})$ is the same as that of $(\pi(B_{\wT t}))_{t\in
[0,\sigma(M)]}$, where $B$ is a Brownian bridge on $\wG$ of length
$\wT \sigma(M)$ from the unit element to $[H_{a_1},H_{b_1}]\ldots
[H_{a_g},H_{b_g}]z$.
\end{lemma}

\pf Let $0\leq t_1 \leq \ldots \leq t_n \leq \sigma(M)$ be $n$ real
numbers. Let $\phi:G^n \lra \R$ and $\psi : G^{2g} \lra \R$ be two
continuous functions. Let us compute the quantity
$\E_{P_{T,z}}[\psi(H_{a_1},H_{b_1},\ldots,H_{a_g},H_{b_g})
\phi(X_{t_1},\ldots,X_{t_n})]$. It is  equal to 

\begin{eqnarray*}
&&\hskip -1cm \frac{1}{Z_{T,z}} \int_{G^{2g+n}}
\psi(a_1,b_1,\ldots,a_g,b_g) \phi(x_1,\ldots,x_n) \\ 
&& \sum_{\pi(\tilde x_1)=x_1,\ldots,\pi(\tilde x_n)=x_n} \tilde p_{T
t_1} (\tilde x_1) \tilde p_{T (t_2-t_1)} (\tilde x_2 \tilde x_1^{-1})
\ldots \tilde p_{T(t_n - t_{n-1})}(\tilde x_n \tilde x_{n-1}^{-1}) \\
&& \hskip 3 cm \tilde p_{\wT (\sigma(M)-t_n)}([\widetilde{a_1,b_1}]
\ldots [\widetilde{a_g,b_g}]\tilde x_n^{-1} z ) \; dx_1 \ldots dx_n \;
da_1 db_1 \ldots da_g db_g \\ 
&=& \hskip 0 cm \frac{1}{Z_{T,z}} \int_{G^{2g}}
\psi(a_1,b_1,\ldots,a_g,b_g) \left[ \frac{1}{\tilde p_{\wT
\sigma(M)}([\widetilde{a_1,b_1}] \ldots [\widetilde{a_g,b_g}] z)}
\int_{\wG^n} \phi(\pi(\tilde x_1),\ldots,\pi(\tilde x_n))\right. \\
&&\hskip -.8cm \left.  \tilde p_{\wT t_1} (\tilde x_1) \tilde p_{\wT
(t_2-t_1)} (\tilde x_2 \tilde x_1^{-1}) \ldots \tilde p_{\wT(t_n -
t_{n-1})}(\tilde x_n \tilde x_{n-1}^{-1}) \tilde p_{\wT
\sigma(M)}([\widetilde{a_1,b_1}] \ldots [\widetilde{a_g,b_g}] z \tilde
x_n^{-1}) \; d\tilde x_1 \ldots d\tilde x_n \right] \\ 
&& \hskip 2.5 cm \tilde p_{\wT \sigma(M)}([\widetilde{a_1,b_1}] \ldots
[\widetilde{a_g,b_g}] z) \; da_1 db_1 \ldots da_g db_g.
\end{eqnarray*}
For all $\tilde y \in \wG$, let $(B^{\tilde y}_t)_{t\in
[0,\wT\sigma(M)]}$ denote a Brownian bridge on $\wG$ of length
$\wT\sigma(M)$ starting at the unit element and finishing at $\tilde
y$. Then the expression between the brackets is exactly
$\E[\phi(\pi(B^{\tilde y}_{\wT t_1}),\ldots, \pi(B^{\tilde y}_{\wT t_n}))]$
with $\tilde y=[\widetilde{a_1,b_1}] \ldots [\widetilde{a_g,b_g}]z$. The
result follows. \qed

Let $\M_c(PM,G)$, or simply $\M_c$, denote the subset of $\M(PM,G)$
consisting of those multiplicative functions $f:PM \lra G$ such that
the mapping from $[0,\sigma(M)] \cap \Q$ to $G$ which sends $t$ to
$f(l_t)$ is uniformly continuous. Observe that $\M_c$ belongs to the
cylinder $\sigma$-field $\C$. As a corollary of the Lemma above, we have
\be \label{as}
\forall T>0, \forall z\in \Pi , \; P_{T,z}(\M_c)=1.
\ee

We can now define two random variables on $(\M(PM,G),\C)$. 

\begin{definition} 1. Let $(X'_t)_{t\in [0,\sigma(M)]}$ be the unique
continuous extension of $(X_t)_{t\in [0,\sigma(M)]\cap \Q}$. Let
$(\widetilde X'_t)_{t\in [0,\sigma(M)]}$ be the lift of $(X'_t)_{t\in
[0,\sigma(M)]}$ to $\wG$, starting at the unit element. Set 
\be
\oo=\widetilde X'_1 \left([\widetilde{H_{a_1},H_{b_1}}] \ldots
[\widetilde{H_{a_g},H_{b_g}}]\right)^{-1}.
\ee

2. Set 
\be 
\tau= \limsup_{t \downarrow 0, t\in \Q}  \frac{d_G(1,X_t)}{\sqrt{2t \log |\log
t|}}.
\ee 
\end{definition}

The random variable $\oo$ is well defined on $\M_c$, hence, by
(\ref{as}), $P_{T,z}-$almost surely for all $T>0$ and $z\in \Pi$. The
variable $\tau$ is well-defined everywhere, possibly equal to
$+\infty$. 

The next result implies Theorem \ref{singularite}.

\begin{proposition} There exists a real positive constant $C$, which
depends only on $G$, such that
$$\forall T>0, \forall z\in \Pi, \; P_{T,z}[\{\oo=z\} \cap \{\tau= C
\sqrt{T}\}]=1.$$ 
\end{proposition}

\pf It follows immediately from Lemma \ref{bridge} that $\oo=z$
$P_{T,z}-$ almost surely.

Let us now fix $T>0$ and $z\in \Pi$ and prove that $\tau = C\sqrt{T}$
$P_{T,z}$- almost surely for some constant $C$.

By Levy's iterated logarithm law (see \cite{McKean}), there exists a
constant $C>0$ such 
that the Brownian motion $(W_t)_{t\geq 0}$ on $\wG$ started from $1$
satisfies almost surely
\be \label{ill}
\limsup_{t\downarrow 0}
\frac{d_\wG(1,W_t)}{\sqrt{2t\log|\log t|}} = C, 
\ee
where $\wG$ is endowed with the metric such that $\pi:\wG \lra G$ is a
local isometry. We denote by $d_\wG$ the corresponding distance.

For all $\tilde y \in \wG$, let $(B^{\tilde y}_t)_{t\in
[0,\wT\sigma(M)]}$ be the Brownian bridge defined in Lemma
\ref{bridge}. For all $\tilde y\in \wG$ and up to any time
$s<\wT\sigma(M)$, the law of $(B_t^{\tilde y})_{t\in [0,s]}$ is
absolutely continuous with respect to that of $(W_t)_{t\in
[0,s]}$. Hence, $B^{\tilde y}$ satisfies the iterated logarithm law
(\ref{ill}).

Now, for all $\tilde x \in \wG$, $d_\wG(1,\tilde x)= d_G(1,\pi(\tilde
x))$. In particular, using the fact that $\wT = (\deg
\rho)^{-\frac{2}{d}} T$,
\be
\limsup_{t\downarrow 0, t\in \Q} \frac{d_G(1,B^{\tilde y}_{\wT
t})}{\sqrt{2t\log|\log t|}} = C\sqrt{T}.  
\ee 
Finally, by Lemma \ref{bridge}, given
$H_{a_1},H_{b_1},\ldots,H_{a_g},H_{b_g}$, $(X_t)_{t \in
[0,\sigma(M)]}$ has the law of a Brownian bridge $(B^{\tilde y}_{\wT
t})_{t\in [0,\sigma(M)]}$ for some $\tilde y$. The result
follows. \qed

\section{Combinatorial computation of the partition functions}

Let $T>0$ and $z\in \Pi$ be fixed. Let $\G$ be a graph on $M$. The
fact that the number $Z^\G_{T,z}$ does not depend on $\G$ is obtained
in \cite{Levy_AMS} as a consequence of a rather tedious approximation
procedure. On the other hand, it is explained in a very convincing, if
not very rigorous, way in \cite{Witten}. In this section, we show
that it is possible to compute $Z^\G_{T,z}$ in a combinatorial way, by
using among other tools the formalism of fat graphs.

\begin{theorem} \label{Z}
Let $g\geq 0$ denote the genus of $M$. Let $\G$ be a graph on $M$. Then, for
each $T>0$ and each $z\in \Pi$, one has 
$$Z^\G_{T,z}=\int_{G^{2g}} \tilde p_{T \tilde \sigma(M)}([\widetilde{a_1,b_1}]
\ldots [\widetilde{a_g,b_g}] z) \; da_1 db_1 \ldots da_g db_g. $$
\end{theorem}

\subsection{Reduction to the case of a graph with a
single face}
\label{subsection reduction}

Let $\G=(\V,\E,\F)$ be a graph on $M$. We do not assume that it is
simple. Let us consider the dual combinatorial graph\footnote{We
consider in fact the combinatorial graph underlying the dual fat graph
to the fat graph induced by $\G$.} of $\G$. It is a pair $\G'=(V,E)$
of finite sets, namely $V=\F$, $E=\E$, endowed with two mappings
$s,t:E \lra V$, respectively defined by $s(e)=L(e)$ and
$t(e)=L(e^{-1})$ (see Section \ref{sect: graphs} for the definition of
$L$). This graph $\G'$ is clearly connected.

A {\it subtree} of $\G'$ is a connected subset $\T \subset E$ stable
by inversion and containing no cycle. Our main tool will be a {\it
spanning tree} of $\G'$, that is, a subtree which is maximal for the
inclusion. Such a subtree satisfies $s(\T)=t(\T)=V$.

The following properties are elementary and we leave their proof to the reader.

\begin{lemma} \label{faits}
Let $\T\subset E$ be a subtree of $\G'$.\\ 
1. The set $\bigcup_{e\in \E\backslash \T} e$ is connected.\\ 
2. The set of endpoints of $\E\backslash \T$ is $\V$.\\ 
3. If $\T$ is a spanning tree, the the set $M \backslash \bigcup_{e\in
\E\backslash \T} e$ is connected.
\end{lemma}

\begin{proposition} \label{single face}
Let $\G=(\V,\E,\F)$ be a graph. Let $\G'=(V,E)$ be
the dual combinatorial graph of $\G$. Let $\T \subset E$ be a spanning
tree of $\G'$. Then $\G_\T=(\V,\E\backslash \T,\{M\})$ is a graph with
a single face and $Z^\G_{T,z}=Z^{\G_\T}_{T,z}$.
\end{proposition}

\pf We proceed by induction on the cardinal of $\T$. If
$\T=\varnothing$, then $\G'$ has only one vertex, so that $\G$ has a
single face and the result is true.

Assume that the result has been proved under the assumption $|\T| \leq
n-1$ for some integer $n\geq 1$. Assume $\G$ and $\T$ are given with
$|\T|=n$. Let $F_1$ be a leaf of $\T$, that is, an element of $V$ such
that $|\{e\in E: s(e)=F_1\}|=|\{e\in E: t(e)=F_1\}|=1$. Let $e\in \T$
be the edge such that $t(e)=F_1$. Set $s(e)=F_2$. Since $\T$ is a
tree, $F_1 \neq F_2$.

Set $\E_e=\E\backslash \{e, e^{-1}\}$ and $\F_e=(\F\backslash
\{F_1,F_2\}) \cup \{F_1 \cup F_2\}$. Then $\G_e=(\V,\E_e,\F_e)$ is
still a graph. The fact that the set of vertices is $\V$ comes from
the second assertion of Lemma \ref{faits}. Moreover, $\T_e=\T\backslash
\{e,e^{-1}\} \subset \E_e$ is a spanning tree of $\G'_e$ and
$|\T_e|=n-2 \leq n-1$.

Now, $\G_e \leq \G$, so that, by the proof of the invariance under
subdivision of the discrete measures, more precisely by (\ref{case E2}),
$Z^{\G_e}_{T,z}=Z^\G_{T,z}$. The result follows by induction. \qed

\subsection{Fat graphs with a single face}
\label{fat graphs}

In order to reduce further the problem, we introduce more carefully
the structure of fat graph. The reader may consult \cite{Lando_Zvonkin} for
further details and also M. Imbert's paper \cite{Imbert} from which
the strategy of our proof is inspired.

Let $\E$ be a set of cardinal $2a$. A structure of fat graph on $\E$
is the data of two permutations $\sigma$ and $\alpha$ of $\E$ such
that $\alpha$ is a fixed-point free involution\footnote{The notation
$\sigma$ is standard and we keep it. There should be no confusion with
the volume $2$-form on $M$.}. 

If $\G=(\V,\E,\F)$ is a graph on $M$, the structure of fat graph on
$\E$ induced by $\G$ is given by setting, for all $e\in \E$,
$\alpha(e)=e^{-1}$ and defining, for each $e\in \E$, $\sigma(e)$ as
the incoming edge at $\overline{e}$ which follows immediately $e$ in
the cyclic order induced by the orientation of $M$. 

The vertices, edges, faces of the fat graph $\Gamma=(\E;
\sigma,\alpha)$ are respectively the cycles of the permutations
$\sigma, \alpha, \alpha \sigma^{-1}$. We denote the last permutation
by $\varphi$, so that $\sigma \alpha \varphi = 1$. If $e\in \E$, the
finishing (resp. starting) point $\overline{e}$ (resp. $\underline{e}$) of
$e$ is defined as the cycle of $\sigma$ containing $e$
(resp. $\alpha(e)$). We use the notation $e^{-1}=\alpha(e)$.

The number of vertices, edges, faces, are denoted respectively by
$s,a,f$. The genus of the graph is the number $g$
defined\footnote{Here again, the conflict of notation with the generic
element of $G$ should not lead to any ambiguity.} by Euler's
relation $s-a+f=2-2g$. If $\G$ is a graph on $M$, it induces on $\E$ a
structure of fat graph with genus equal to that of $M$.

Let $\Gamma=(\E;\sigma,\alpha)$ be a fat graph with a single face,
that is, such that $\varphi=\alpha \sigma^{-1}$ is a cyclic permutation
of $\E$. Consider the adjoint action $\Ad$ of $\wG$ on itself. We
associate to $\Gamma$ a probability measure on $\wG/\Ad$ as follows.

Assume $\E=\{e_1,\ldots,e_a,\alpha(e_1),\ldots,\alpha(e_a)\}$. Write
$\varphi=(\alpha^{i_1}(e_{n_1}),\ldots,\alpha^{i_{2a}}(e_{n_{2a}}))$, with
$i_k \in \{0,1\}$ for $k=1\ldots 2a$. Then one can define a mapping
$h_\Gamma : G^a \lra \wG/\Ad$ by setting, for each
$g=(g_1,\ldots,g_a)$, $h_\Gamma(g)=\tilde g_{n_{2a}}^{\epsilon_{2a}}
\ldots \tilde g_{n_1}^{\epsilon_1}$, where $\tilde g=(\tilde
g_1,\ldots,\tilde g_r)$ is an arbitrary lift of $g$ and, for each
$k=1\ldots 2a$, $\epsilon_k=1-2 i_k$. Since each edge appears twice
in $\varphi$, once with each orientation, this definition does not depend
on the choice of $\tilde g$.

The mapping $h_\Gamma$ depends on the choice of $e_1,\ldots,e_a \in
\E$ but the image of the Haar measure on $G^a$ by $h_\Gamma$ does
not. We associate to $\Gamma$ the probability measure
$\nu_\Gamma=(h_\Gamma)_* (dg^{\otimes r})$. We think of $\nu_\Gamma$
as a measure on $\wG/\Ad$ or a measure on $\wG$ invariant by adjunction.
The following lemma is a straightforward consequence of the definition
of $\nu_\Gamma$.

\begin{lemma} \label{Z nu}
Let $\G$ be a graph on $M$ with a single face. Let
$\Gamma$ be the fat graph induced by $\G$. Then 
\be 
Z^\G_{T,z}=  \int_\wG \tilde
p_{T\tilde \sigma(M)}(\tilde g z) \; d\nu_\Gamma(\tilde g).  
\ee
\end{lemma}

Let us recall two classical operations on fat graphs, namely the
contraction of an edge, or Whitehead's move, and the cut-and-paste
operation. 

\begin{definition} Let $\Gamma=(\E;\sigma,\alpha)$ be a fat
graph. \\
1. {\sl Whitehead's move} --  Let $e\in \E$ be given such that
$\underline{e} \neq 
\overline{e}$. Set $\E'=\E\backslash \{e,\alpha(e)\}$ and define
$\alpha'=\alpha_{|\E'}$. Decompose $\sigma$ in a product of commuting
cycles and write $\sigma=(e,e_1,\ldots,e_k)
(\alpha(e),e_{k+1},\ldots,e_l)\sigma_0$ where $\sigma_0$ is the product of the
cycles which contain neither $e$ nor $\alpha(e)$. Set
$\sigma'=(e_1,\ldots,e_k,e_{k+1},\ldots, e_l)\sigma_0 $. The fat graph
$(\E';\sigma',\alpha')$ is by definition the result of the contraction
of the edge $e$ in $\Gamma$. It is denoted by $W_e(\Gamma)$.\\
2. {\sl Cut and paste} -- Assume that $\Gamma$ has a single face. Let
$e \in \E$ be 
given. Write $\varphi=\alpha \sigma^{-1}$ as
$\varphi=(e,e_1,\ldots,e_r,d,e^{-1},e_{r+1},\ldots,e_s),$ where we
have emphasized $d=\varphi^{-1}(e^{-1})$. Set
$\varphi'=(e,d,e_1,\ldots,e_r,e^{-1},e_{r+1},\ldots,e_s)$ and
$\sigma'=(\varphi')^{-1} \alpha$. Then the fat graph
$(\E;\sigma',\alpha)$ is by definition the result of the cut-and-paste
operation along $e$. It is denoted by $K_e(\Gamma)$.
\end{definition}

The following properties are classical (see \cite{Imbert}).

\begin{lemma} Let $\Gamma$ be a fat graph. Let $e,e'$ be two edges
of $\Gamma$.\\ 
1. Assume that $\underline{e} \neq \overline{e}$. Then
$W_e(\Gamma)$ is a fat graph with the same number of faces, same
genus, and one less vertex as $\Gamma$.\\ 
2. Whitehead's moves commute : if $W_e(W_{e'}(\Gamma))$ and
$W_{e'}(W_e(\Gamma))$ are both defined, they are equal.\\
3. Assume that $\Gamma$ has one single face and one single
vertex. Then $K_e(\Gamma)$ still has one single face and
one single vertex.
\end{lemma}

Let us describe how the measure $\nu$ is transformed by these
operations.

\begin{proposition} \label{invariance nu} Let $\Gamma=(\E;\sigma,\alpha)$
be a fat graph with a single face. Pick $e\in \E$.\\ 
1. One has $\nu_\Gamma =\nu_{K_e(\Gamma)}$. \\ 
2. If $e$ has distinct endpoints, then $\nu_\Gamma=\nu_{W_e(\Gamma)}$.\\
\end{proposition}

\pf 1. Write $\E=\{e_1,\ldots,e_a,\alpha(e_1),\ldots,\alpha(e_a)\}$
with $e_1=e$ and $e_2=\sigma(e_1)=\varphi^{-1}(e^{-1})$. We construct a
change of variables $k=(k_1,\ldots,k_a):G^a \lra G^a$ as
follows. Define $k_1=g_2^{-1}g_1$ and $k_i=g_i$ for $i=2\ldots
a$. This change of variables satisfies the relation
$h_\Gamma(g)=h_{K_e(\Gamma)}(k)$ and preserves the Haar measure. The
result follows. 

2. Consider the other change of variables $w=(w_1,\ldots,w_a):G^a \lra
G^a$ defined by setting $w_1=g_1$ and, for each $i=2\ldots a$,
$$w_i = \left\{ \begin{array}{ll} g_i g_1 & {\rm if} \;\; \underline{e_i}=\overline{e} \;\; {\rm and} \;\; \overline{e_i}\neq
\overline{e}, \\
g_1^{-1} g_i & {\rm if} \;\;
\underline{e_i}\neq \overline{e} \;\; {\rm and} \;\; \overline{e_i}=
\overline{e}, \\
g_1^{-1} g_i g_1 & {\rm if} \;\;
\underline{e_i}= \overline{e} \;\; {\rm and} \;\; \overline{e_i}=
\overline{e}. \end{array} \right.$$

This change of variables satisfies the relation
$h_\Gamma(g_1,\ldots,g_a)=h_{W_e(\Gamma)}(w_1,\ldots,w_a)$ and
preserves the Haar measure. This finishes the proof. \qed

\begin{definition} \label{standard}
Let $\Gamma=(\E;\sigma,\alpha)$ be a fat graph of
genus $g$ with a single face and a single vertex. Let $m$ be an
integer such that $0 \leq m \leq g$. We say that $\G$ is {\sl standard
of order} $m$ if it is possible to label the elements of $\E$ in such
a way that
$\varphi=(a_1,b_1,a_1^{-1},b_1^{-1},\ldots,a_m,b_m,a_m^{-1},b_m^{-1},
e_{4m+1},\ldots,e_{4g})$.
\end{definition}

If $\Gamma$ is standard of order $g$, it is very easy to write down
the measure $\nu_\Gamma$. The next result allows us to extend this
observation to the general case.

\begin{proposition}  \label{inc standard}
Let $\Gamma=(\E;\sigma,\alpha)$ be a fat graph of genus $g$ with a
single face and a single vertex. Assume that $g>0$ and $\Gamma$ is
standard of order $m$ for some $m<g$. Then there exists a fat graph
$\Gamma'$ standard of order $m+1$ and such that $\nu_\Gamma=
\nu_{\Gamma'}$.
\end{proposition}

\pf By Euler's relation, $\Gamma$ has $2g$ edges, so that $|\E|=4g\geq
4$. Write
$$\varphi=(a_1,b_1,a_1^{-1},b_1^{-1},\ldots,
a_m,b_m,a_m^{-1},b_m^{-1},e,e_1,\ldots,e_k,e^{-1},e_{k+1},\ldots,e_{k+l})$$
with $k,l\geq 0$ and $k+l=4(g-m)-2$. From now on, we abbreviate
$a_1,\ldots,b_m^{-1}$ by $S_m$.

We claim that $k$ is positive. Otherwise, $\sigma=\varphi^{-1} \alpha$
would fix $e$, in contradiction with the fact the $\Gamma$ has a
single vertex.

For the same reason, it is not possible that
$\{e_1,\ldots,e_k\}$ be stable by the permutation $\alpha$. Otherwise,
$\sigma$ would stabilize $\{e,e_1,\ldots,e_k\}$ which is a proper
subset of $\E$ since it does not contain $e^{-1}$. Hence, $\varphi$
can be written, with a new labeling of the edges,
$$\varphi=(S_m,e,e_1,\ldots,e_r,f,e_{r+1},\ldots,e_{s},e^{-1},e_{s+1},
\ldots,e_{t},f^{-1},e_{t+1},\ldots,e_{u})$$ 
with $0\leq r \leq s \leq t \leq u$. Consider now
$$\Gamma'= K_{f^{-1}}^{t} \circ K_e^{t-r} \circ K_f^{t-s} (\Gamma).$$
Then, if $\Gamma'=(\E;\sigma',\alpha)$, with the same labeling of
$\E$, one has 
$$\varphi'=\alpha (\sigma')^{-1}=(S_m,e,f,e^{-1},f^{-1},
e_{s+1}\ldots,e_t,e_{r+1},\ldots,e_s,e_1,\ldots,e_r,e_{t+1},\ldots,e_u).$$
Thus, $\Gamma'$ is standard of order $m+1$ and, by Proposition
\ref{invariance nu}, it satisfies $\nu_{\Gamma'}=\nu_\Gamma$. The result is
proved. \qed

{\noindent \textbf{Proof of Theorem \ref{Z} -- }} Let $\G$ be a graph
on $M$. Pick $T>0$ and $z\in \Pi$. Let us compute $Z^\G_{T,z}$.

By Proposition \ref{single face}, we may assume that $\G$ has a
single face. Let $\Gamma=(\E;\sigma,\alpha)$ be the fat graph induced
by $\G$. By Lemma \ref{Z nu}, it is enough to compute $\nu_\Gamma$.

Assume that $\Gamma$ has at least two vertices. By applying Whitehead's
moves along the edges of a spanning tree of $\Gamma$, we transform
$\Gamma$ into a fat graph with one single vertex. According to
Proposition \ref{invariance nu}, this leaves the measure $\nu_\Gamma$
unchanged. Thus, we may assume that $\Gamma$ has one single face and
one single vertex.

If the genus of $\Gamma$ is $0$, that is, if $M$ is a sphere, then
$\Gamma$ has no edges and the measure $\nu_\Gamma$ is the Dirac mass
at the unit element of $\wG$. 

If the genus of $\Gamma$ is positive, let $m\geq 0$ be the greatest
integer such that $\Gamma$ is standard of order $m$. If $m=g$, then
$\nu_\Gamma$ is the image on $\wG$ of the Haar measure on $G^{2g}$ by
the mapping $(a_1,b_1,\ldots,a_g,b_g) \mapsto
[\widetilde{a_1,b_1}]\ldots [\widetilde{a_g,b_g}]$.  If $m<g$, then by
induction on $g-m$, Proposition \ref{inc standard} implies that
$\nu_\Gamma$ is the same as if $m=g$.

In conclusion, $\nu_\Gamma$ is always the image on $\wG$ of the Haar
measure on $G^{2g}$ by the mapping $(a_1,b_1,\ldots,a_g,b_g) \mapsto
[\widetilde{a_1,b_1}]\ldots [\widetilde{a_g,b_g}]$. By Lemma \ref{Z
nu}, this proves the result. \qed

\section{Appendix: The new discrete theory as a singular covering of
the old one} 

In this appendix, we study the structure of the vertical arrow of
Diagram (\ref{lift H}). We prove that it is a covering outside a
closed singular set of codimension one. First, let us recall a nice
description of the base space $G^{\E^+}/\J_{\G}$ which is explained in
\cite{Durhuus} and \cite{Levy_JGP}. In what follows, $\G=(\V,\E,\F)$
is a graph, that we do not assume to be simple. We assume an
orientation $\E^+=\{e_{1},\ldots,e_{r}\}$ has been chosen.

%If $c=e_{i_1}^{\epsilon_1}\ldots e_{i_n}^{\epsilon_n}$ is a path in
%$\G$, with $\epsilon_1,\ldots,\epsilon_n=\pm 1$, then we define the
%discrete holonomy mappings $h_c:G^{\E^+} \lra G$ and $\tilde h_c: \EC
%\lra \wG$ by setting respectively
%$h_c(g_1,\ldots,g_r)=g_{i_n}^{\epsilon_n} \ldots g_{i_1}^{\epsilon_1}$
%and $\tilde h_c(\tilde g)= \tilde g({e_{i_n}^{\epsilon_n}}) \ldots
%\tilde g(e_{i_1}^{\epsilon_1})$.

Let $T\subset\E$ be a spanning tree of $\G$. By this we mean, as at
the beginning of Section \ref{subsection reduction}, that $T$ is a
connected subset of $\E$ stable by inversion, that no simple loop can
be made by concatenating edges of $T$ and that $T$ is maximal for
inclusion with these properties. In particular, the set of endpoints
of the edges of $T$ is $\V$ itself. Hence, if $v,w\in\V$, there is a
unique injective path from $v$ to $w$ within $T$. We denote this path
by $[v,w]$. Finally, let us choose a vertex $r$ that we call the 
root. For each edge $e\in \E$, define the loop $\lambda_{e}$ based at
$r$ by $\lambda_{e}=[r,\underline{e}] e [\overline{e},r]$.

\begin{proposition} \label{standard ususel}
The mapping $G^{\E^+} \lra G^{\E^+ \backslash
T^+}/G$ which sends $g=(g_1,\ldots,g_r)$ to the orbit of
$(h_{\lambda_e}(g))_{e\in\E^+\backslash T^+}$ under the action of $G$ by
diagonal conjugation on $G^{\E^+ \backslash T^+}$ induces a
homeomorphism
$$G^{\E^+}/\J_\G \lra G^{\E^+ \backslash T^+} /G.$$
\end{proposition}

This proposition says that any configuration is gauge-equivalent to a
configuration which is equal to $1$ on the edges of
$T$\footnote{Considering configurations which vanish on the edges of
$T$ is the discrete analog to putting connections in axial gauge with
respect to some coordinate system.} and also that two such
configurations are equivalent if and only if they differ by
simultaneous conjugation by some element of $G$.

Using this homeomorphism, it is possible to describe in a simple way
most of the space $\EC$. In order to explain what {\it most} means, we
use the following result.

\begin{proposition} Let $x$ be an element of $G$. Let $\tilde x \in \wG$ be an
element of $\pi^{-1}(x)$. The subgroup $\{z\in \Pi \; | \; z\tilde x
\in \Ad(\wG) \tilde x \}$ of $\Pi$ does not depend on the choice of
$\tilde x$. Let $S\subset G$ be the set of those $x$ for which this
subgroup is not reduced to $\{1\}$. Then $S$ is stable by conjugation,
closed and of codimension 1 in $G$. More precisely, it is contained in
the smooth image of a manifold of dimension at most $\dim G -1$.
\end{proposition}

\pf Only the two last assertions are not elementary. According to a
general structure theorem (\cite[Theorem V.8.1]{Broecker_tom_Dieck}),
$\wG$ is isomorphic to $\R^m\times K$, where $m\geq 0$ and $K$ is
simply connected and semi-simple. Define $\Pi_K=\{c \in K \; | \;
(1,c) \in \Pi\}$. Let $\tilde x, \tilde y \in \wG$ and $z \in \Pi$ be
such that $\tilde y \tilde x \tilde y^{-1}= z\tilde x$. Then,
decomposing this identity according to $\wG \simeq \R^m \times K$
shows that $\tilde x$ belongs to $\{1\} \times S_K$, where $S_K=\{ k
\in K \; | \; \exists c \in \Pi_K, ck \in \Ad(K)k\}$. In fact, the
equality $S=\pi(\{1\} \times S_K)$ holds.

Since $\Pi_K \subset Z(K)$ is finite and $K$ is compact, $S_K$ and
hence $S$ are closed. Let $T$ be a maximal torus of $K$. We claim that
$S_K \cap T$ is a finite union of cosets of proper subgroups of $T$. 

Indeed, let $t\in S_K \cap T$. There exists $c \in \Pi_K$, $c\neq 1$,
such that $t$ and $ct$ are conjugate. Let $W$ be the Weyl group of
$T$. Since both $t$ and $zt$ belong to $T$, there exists $w\in W$,
$w\neq \id$, such that $w(t)=ct$. So, we have proved that 
$$S_K \cap T \subset \bigcup_{\stackrel{\scriptstyle c\in
\Pi_K-\{1\}}{w\in W-\{\id\}}} T_{c,w},$$  
where $T_{c,w}=\{ t \in T \; | \; w(t)=ct\}$. 

Let $c,w$ be given as above. If $T_{c,w}$ is not empty, it is a coset
of the closed subgroup $\{ t\in T \; | \; w(t)t^{-1}=1\}$. Since
$w\neq \id$, this subgroup is a proper subgroup and our claim is
proved. 

Finally, $S_K$ is the union of the images of the mappings $K/T \times
T_{c,w} \lra K$, $(kT,t) \mapsto \Ad(k)t$ and $\dim (K/T \times
T_{c,w}) \leq \dim K -1 \leq \dim G - 1$. \qed 

As usual, we denote by $G/\Ad$ and $\wG/\Ad$ the spaces of conjugacy
classes of $G$ and $\wG$ respectively. We identify $S$ with a subset
of $G/\Ad$ and $\wS=\pi^{-1}(S)$ with a subset of $\wG/\Ad$. 

\begin{example} \label{exemple S} Take $\wG=SU(n)$ and
$\Pi=Z(SU(n))$. An element of $SU(n)$ is in the set $\wS$ if and only
if its spectrum is invariant by multiplication by some $n$-th root of
unity distinct from 1. If the spectrum of a matrix is invariant by
multiplication by a primitive $d$-th root $\zeta$ of unity, with
$d|n$, then its spectrum is of the form $\{\alpha_1, \alpha_1 \zeta,
\ldots, \alpha_1 \zeta^{d-1},\ldots, \alpha_{\frac{n}{d}}, \ldots,
\alpha_{\frac{n}{d}} \zeta^{d-1} \}$, with $\alpha_1 \ldots
\alpha_{\frac{n}{d}}=1$. In a maximal torus of $SU(n)$, of dimension
$n-1$, such matrices form a union of submanifolds of codimension
$n-\frac{n}{d}$. Hence, $\wS$ is of codimension at least
$n-\frac{n}{d}$ in $SU(n)$, where $d$ is the smallest divisor of $n$
greater or equal to 2. In any case, $\codim \wS \geq \frac{n}{2}$.
\end{example}

The main consequence of the definition of $S$ is the following.

\begin{lemma} \label{revetement} The group $\Pi$ acts freely and
properly by translations on $\wG/\Ad - \wS$, and the quotient space of
this action is canonically homeomorphic to $G/\Ad - S$.
\end{lemma}

\pf The translate of a conjugacy class of $\wG$ by a central element
is still a conjugacy class. Hence, $\Pi$ acts by translations on
$\wG/\Ad$. It is elementary to check that the orbits of this action
are in bijective correspondence with the conjugacy classes of
$G$. Moreover, the natural projection $\wG \lra G/\Ad$ induces a
continuous mapping $\Pi\backslash \wG \lra G/\Ad$. Since the adjoint
action commutes to that of $\Pi$, this induces in turn a continuous
bijective mapping $\Pi\backslash \wG/\Ad \lra G/\Ad$. Since the source
space of this mapping is compact, it is a homeomorphism.

The subset $S$ of $G$ has been defined precisely in such a way that
the restriction of the action of $\Pi$ to $\wG/\Ad - \wS$ is free. To
show that it is proper, write as in the last proof $\wG\simeq \R^m
\times K$, with $K$ semi-simple. Observe then that $\wG/\Ad \simeq
\R^m \times K/\Ad$. Then $\Pi$ is a subgroup of $L \times Z(K)$, where
$L$ is some lattice in $\R^m$, for example the projection of $\Pi$ on
the first factor. Now since $Z(K)$ is finite, $L\times Z(K)$ acts
properly on $\R^m \times K/\Ad$. Hence, so does $\Pi$ on
$\wG/\Ad$. \qed

\begin{remark} It is not always true that $\wG/\Ad -
\wS$ is connected. In particular, one cannot say that the projection
of $\wG/\Ad -\wS$ on $G/\Ad - S$ is a Galois covering with
automorphism group $\Pi$. For example, take $G=SO(3)$, $\wG=SU(2)$ and
$\Pi=\{I_2,-I_2\}$.  Then $S$ is the set of matrices of rotations with
angle $\pi$ in $\R^3$ and $\wS$ is the set of $2\times 2$ unitary
matrices with spectrum $\{i,-i\}$. This is the equatorial 2-sphere
of $SU(2)\simeq S^3$ so that $SU(2)- \wS$ is not connected. On the
other hand, take $\wG=SU(3)$ and $\Pi=\{I_3,jI_3,j^2I_3\}$. Then $\wS$
is the set of $3\times 3$ unitary matrices with spectrum
$\{1,j,j^2\}$, which is a smooth submanifold of codimension 2. In this
case, $SU(3) - \wS$ is connected and the projection is a Galois
covering. In fact, according to Example \ref{exemple S}, $\wG/\Ad -
\wS$ is always connected when $\wG=SU(n)$ with $n\geq 3$.
\end{remark} 

Let us define $U\subset G^{\E^+}/\J_\G$ as the open set of configurations
such that the holonomy along the boundary of each face belongs to the
complement of $S$. We shall now analyze the vertical arrow of (\ref{lift H})
restricted to $\pi^{-1}(U)$.

\begin{proposition} The restriction of the mapping $\pi: \EC/\wJ_\G \lra
G^{\E^+}/\J_G$ to $\pi^{-1}(U)$ is a covering of $U$ whose
automorphism group acts transitively on the fibers and is isomorphic
to $\Pi^\F$.  
\end{proposition}

\pf In this proof, the generic element of $\wG/\Ad$ is denoted by
$\tilde\xx$. We denote also by $\pi:\wG/\Ad \lra G/\Ad$ the natural
projection. Let $X \subset G^{\E^+}/\J_G \times (\wG/\Ad)^\F$ be
defined as
$$X=\{ ([g],(\tilde \xx_F)_{F\in \F}) \; | \; \forall F \in \F, h_{\partial
F}([g]) = \pi(\tilde \xx_F)\}.$$ 

Set $X_U= X \cap (\pi^{-1}(U) \times (\wG/\Ad)^\F)$.  Define a mapping
$\kappa : \EC/\wJ_\G \lra X$ by setting $\kappa(\tilde g)= (\pi(\tilde
g), (\tilde h_{\partial F}(\tilde g))_{F\in \F})$. By definition,
$\pi=pr_1 \circ \kappa$. We claim two things which together imply our
result. \\ 
1. The mapping $\kappa$ induces a homeomorphism between
$\pi^{-1}(U)$ and $X_U$.\\ 
2. The projection $pr_1 : X_U \lra U$ is a covering whose automorphism
group acts transitively on the fibers and is isomorphic to $\Pi^\F$. \\
{\it Proof of claim 1 -- } To begin with, $\kappa$ is a continuous
mapping. Let us prove that it is onto. For this, choose $([g],(\tilde
\xx_F)_{F\in\F})$ in $X_U$. Pick $\tilde g\in \EC$ such that
$\pi(\tilde g)=g$. Let $F$ be a face of $\G$. We have $\pi(\tilde
h_{\partial F}(\tilde g))=\pi(\tilde \xx_F)$. Hence, there exists
$z\in \Pi$ such that $\tilde h_{\partial F}(\tilde g)z=\tilde
\xx_F$. Let $e\in\E$ be such that $L(e)=\F$. Replacing $\tilde g(e)$
by $\tilde g(e)z$, we transform $\tilde g$ without changing
$\pi(\tilde g)$ into a configuration such that the class of the
holonomy along the boundary of $F$ is exactly $\tilde \xx_F$. We can
do this for each face successively and we get a configuration $\tilde
g$ such that $\kappa(\tilde g)=([g],(\tilde \xx_F)_{F\in\F})$. 

Let us prove that it is one-to-one. For this, consider again the
spanning tree $T \subset \E$ and the root $r\in\V$. Let $\tilde g$ and
$\tilde g'$ be two configurations such that $\kappa(\tilde
g)=\kappa(\tilde g')$. Let us prove that they are equivalent. First,
each of them is equivalent to a configuration which takes its values
in $\Pi$ on the edges of $T$. Indeed, take $\tilde g$ for example. The
relations $j_r=1$ and $j_{\overline {e}}^{-1} \tilde g(e)
j_{\underline{e}}=1$ for each $e\in T^+$ determine uniquely $j_v$ for
each vertex $v$. By letting $j=((j_v)_{v\in\V},1) \in \wJ_\G$ act on
$\tilde g$, we get a configuration of the desired form. Assume now
that, for each $e\in T$, both $\tilde g(e)$ and $\tilde g'(e)$ belong
to $\Pi$. Since $\kappa(\tilde g)=\kappa(\tilde g')$, it is in
particular true that the configurations $(\pi(\tilde
g(e)))_{e\in\E^+}$ and $(\pi(\tilde g'(e)))_{e\in\E^+}$ of $G^{\E^+}$
are equivalent. Since both take the value 1 on the edges of $T$ and
according to Proposition \ref{standard ususel}, they differ by
simultaneous conjugation by some element of $G$, say $(\pi(\tilde
g'(e)))=\Ad(x) (\pi(\tilde g(e)))$. Choose $\tilde x$ in $\pi^{-1}(x)$
and set $j_v=\tilde x$ for all $v\in \V$. Then, replacing $\tilde g$
by $((j_v),1)\cdot \tilde g$, we may assume that, for each $e\in \E$,
$\pi(\tilde g(e))=\pi(\tilde g'(e))$. In other words, there exists a
function $z:\E \lra \Pi$ such that, for each $e\in \E$, $\tilde
g'(e)=\tilde g(e) z(e)$. In particular, this implies that, for each
face $F$ of $\G$, the relation $\tilde h_{\partial F}(\tilde
g')=\tilde h_{\partial F}(g) \prod_{L(e)=F} z(e)$ holds. On the other
hand, since $\kappa(\tilde g)=\kappa(\tilde g')$, both sides of this
equality are conjugate. But, by the assumption that both $\tilde g$
and $\tilde g'$ belong to $\pi^{-1}(U)$, this imposes the relation
$\prod_{L(e)=\F} z(e)=1$. Hence, $(1,(z(e))_{e\in\E})$ belongs to the
gauge group and transforms $\tilde g$ into $\tilde g'$, which are
henceforth equivalent. Finally, $\kappa$ is a continuous bijection.

We prove now that its inverse is continuous. Pick $([g],(\tilde
\xx_F)_{F\in\F})$ in $X_U$. Choose a small neighbourhood $V$ of
$\bigcup_{e\in\E^+} \{g(e)\}$ in $G$ and a continuous section $\tau: V
\lra \wG$ of $\pi$. This section induces another continuous section
$\upsilon: W \subset G^{\E^+} \lra \EC$ of the natural projection,
defined on a neighbourhood of $g$ by $\upsilon(g')(e)=\tau(g'(e))$ if
$e\in\E^+$ and $\tau(g'(e^{-1}))^{-1}$ if $e^{-1} \in \E^+$. Finally,
let $\lambda : \F \lra \E$ be a section of $L$, that is, the choice of
an edge on the boundary of each face. For $g'$ in a neighbourhood of
$g$ and $(\tilde\xx'_F)_{F\in\F}$ in a neighbourhood of
$(\tilde\xx_F)_{F\in\F}$, define $\psi(g',(\tilde \xx'_F)) \in \EC$ by
setting $\psi(g',(\tilde\xx'_F))(e)=\upsilon(g')(e)$ if $e\notin
\lambda(\F)$ and $\upsilon(g')(e) z_F$ if $e=\lambda(F)$, where $z_F$
is the unique element of $\Pi$ such that $\tilde h_{\partial F}(g) z_F
= \tilde \xx_F$. Then, by Lemma \ref{revetement} $\tilde
h_{\partial_F}(\psi(g',(\tilde \xx'_F)))=\xx'_F$ for $g'$ close enough
to $g$ and $\xx'_F$ close enough to $\xx_F$. This construction
provides us locally with a continuous inverse to $\kappa$, which is
thus a homeomorphism.

{\it Proof of claim 2 -- } To prove this assertion, observe that $U$
is homeomorphic to the subset $\{ ([g],(\xx_F)_{F\in \F}) \; |
\; \forall F \in \F, h_{\partial F}([g]) = \xx_F\} $ of $G^{\E^+}
\times (G/\Ad)^\F$. Now the result is a consequence of Lemma
\ref{revetement}. \qed

\bibliographystyle{plain}
\bibliography{ntb}

\begin{thebibliography}{10}

\bibitem{Broecker_tom_Dieck}
Theodor Br{\"o}cker and Tammo tom Dieck.
\newblock {\em Representations of compact {L}ie groups}.
\newblock Springer-Verlag, New York, 1995.

\bibitem{Driver_lassos}
Bruce~K. Driver.
\newblock Y{M}${}\sb 2$: continuum expectations, lattice convergence, and
  lassos.
\newblock {\em Comm. Math. Phys.}, 123(4):575--616, 1989.

\bibitem{Driver}
Bruce~K. Driver.
\newblock Two-dimensional {E}uclidean quantized {Y}ang-{M}ills fields.
\newblock In {\em Probability models in mathematical physics (Colorado Springs,
  CO, 1990)}, pages 21--36. World Sci. Publishing, Teaneck, NJ, 1991.

\bibitem{Durhuus}
B.~Durhuus.
\newblock On the structure of gauge invariant classical observables in lattice
  gauge theories.
\newblock {\em Lett. Math. Phys.}, 4(6):515--522, 1980.

\bibitem{Fleischhack_B}
Christian Fleischhack and Jerzy Lewandowski.
\newblock Breakdown of the action method in gauge theories.
\newblock {\em Preprint}, math-ph/0111001, 2001.

\bibitem{Imbert}
Michel Imbert.
\newblock Fundamental groups of fat-graphs.
\newblock {\em Note Mat.}, 19(2):257--268 (2001), 1999.

\bibitem{Lando_Zvonkin}
Sergei~K. Lando and Alexander~K. Zvonkin.
\newblock {\em Graphs on surfaces and their applications}, volume 141 of {\em
  Encyclopaedia of Mathematical Sciences}.
\newblock Springer-Verlag, Berlin, 2004.
\newblock With an appendix by Don B. Zagier, Low-Dimensional Topology, II.

\bibitem{Levy_AMS}
Thierry L{\'e}vy.
\newblock Yang-{M}ills measure on compact surfaces.
\newblock {\em Mem. Amer. Math. Soc.}, 166(790):xiv+122, 2003.

\bibitem{Levy_JGP}
Thierry L{\'e}vy.
\newblock Wilson loops in the light of spin networks.
\newblock {\em J. Geom. Phys.}, 52(4):382--397, 2004.

\bibitem{Levy_Norris}
Thierry L{\'e}vy and James~R. Norris.
\newblock A large deviation principle for the {Y}ang-{M}ills measure.
\newblock {\em Preprint}, math-ph/0406027, 2004.

\bibitem{McKean}
H.~P. McKean, Jr.
\newblock {\em Stochastic integrals}.
\newblock Probability and Mathematical Statistics, No. 5. Academic Press, New
  York, 1969.

\bibitem{Migdal}
A.~A. Migdal.
\newblock Recursion equations in gauge field theories.
\newblock {\em Sov. Phys. JETP}, 42(3):413--418, 1975.

\bibitem{Morita}
Shigeyuki Morita.
\newblock {\em Geometry of differential forms}, volume 201 of {\em Translations
  of Mathematical Monographs}.
\newblock American Mathematical Society, Providence, RI, 2001.

\bibitem{Rao}
M.~M. Rao.
\newblock Projective limits of probability spaces.
\newblock {\em J. Multivariate Anal.}, 1(1):28--57, 1971.

\bibitem{Sengupta_AMS}
Ambar Sengupta.
\newblock Gauge theory on compact surfaces.
\newblock {\em Mem. Amer. Math. Soc.}, 126(600):viii+85, 1997.

\bibitem{Steenrod}
Norman Steenrod.
\newblock {\em The topology of fibre bundles}.
\newblock Princeton Landmarks in Mathematics. Princeton University Press,
  Princeton, NJ, 1999.

\bibitem{Uhlenbeck}
Karen~K. Uhlenbeck.
\newblock Connections with {$L\sp{p}$} bounds on curvature.
\newblock {\em Comm. Math. Phys.}, 83(1):31--42, 1982.

\bibitem{Wehrheim}
Katrin Wehrheim.
\newblock {\em Uhlenbeck compactness}.
\newblock EMS Series of Lectures in Mathematics. European Mathematical Society
  (EMS), Z\"urich, 2004.

\bibitem{Witten}
Edward Witten.
\newblock On quantum gauge theories in two dimensions.
\newblock {\em Comm. Math. Phys.}, 141(1):153--209, 1991.

\end{thebibliography}

\end{document}